\documentclass[11pt]{article}
\usepackage[T1]{fontenc}
\usepackage[latin9]{inputenc}
\usepackage{amsmath,amssymb,amsthm,relsize,enumitem,stmaryrd}
\theoremstyle{theorem}

\newtheorem{proposition}{Proposition}
\newtheorem*{proposition*}{Proposition}
\newtheorem{lemma}{Lemma}
\newtheorem{remark}{Remark}
\newtheorem{definition}{Definition}
\newtheorem{corollary}{Corollary}

\title{A Kripke-Lewis semantics for belief update and belief revision}
\author{Giacomo Bonanno\\
{\small University of California, Davis, USA}\\
{\small gfbonanno@ucdavis.edu}
}
\date{\today}

\begin{document}
\maketitle
\begin{abstract}
We provide a new characterization of both belief update and belief revision in terms of a Kripke-Lewis semantics. We consider frames consisting of a set of states, a Kripke belief relation and a Lewis selection function. Adding a valuation to a frame yields a model. Given a model and a state, we identify the initial belief set $K$ with the set of formulas that are believed at that state and  we identify either the updated belief set $K\diamond\phi$ or the revised belief set $K\ast\phi$ (prompted by the input represented by formula $\phi$) as the set of formulas that are the consequent of conditionals that (1) are believed at that state and (2) have $\phi$ as antecedent. We show that this class of models characterizes both the Katsuno-Mendelzon (KM) belief update functions and the AGM belief revision functions, in the following sense: (1) each model gives rise to a partial belief function that can be completed into a full KM/AGM update/revision function, and (2) for every KM/AGM update/revision function there is a model whose associated belief function coincides with it. The difference between update and revision can be reduced to two semantic properties that appear in a stronger form in revision relative to update, thus confirming the finding by Peppas et al. (1996) that, "for a fixed theory $K$, revising $K$ is much the same as updating $K$".  \\[4pt]
\end{abstract}
\section{Introduction}
The notion of belief revision is normally associated with the seminal contribution by Alchourr\'{o}n, G\"{a}rdenfors and Makinson (henceforth AGM) in \cite{AGM85}, while the notion of belief update was formally introduced by Katsuno and Melndelson (henceforth KM) in \cite{KatMen91}. The difference between the two notions is usually explained in terms of a static vs dynamic world: revision occurs when an agent changes his beliefs in response to new information about a static world, while update occurs when an agent modifies his beliefs to keep them up-to-date with an evolving world. Peppas et al. in \cite{Pepetal96} reconsidered the distinction between revision and update by comparing the two processes in terms of their construction from pre-orders on possible worlds (\cite{Gro88,KatMen91}) and showed that, "essentially the results of revision can be duplicated using the construction for update" thus concluding that, for a \emph{fixed} initial belief set $K$, "revising $K$ is much the same as updating $K$" (\cite[p. 95]{Pepetal96}. We reach the same conclusion, but from a different route, by using a Kripke-Lewis semantics that establishes a connection between both revision and update, on one hand, and  belief in conditionals, on the other hand.
\par
We consider frames consisting of a set of states, a Kripke belief relation and a Lewis selection function. Adding a valuation to a frame yields a model. Given a model and a state $s$, we identify the initial belief set $K$ with the set of formulas that are believed at state $s$ and interpret the revised/updated belief set (prompted by the input represented by formula $\phi$) as the set of formulas that are the consequent of conditionals that are believed at state $s$ and have $\phi$ as antecedent; that is, $\psi\in K\ast\phi$ (in the case of revision) or $\psi\in K\diamond\phi$ (in the case of update) if and only if at state $s$ the agent believes that "if $\phi$ is (or were) the case, then $\psi$ is (or would be) the case".  We show that this class of models characterizes both the AGM belief revision functions and the Katsuno-Mendelson (KM) belief update functions, in the following sense: (1) each model (in the appropriate class) gives rise to a partial belief function that can be completed into a full AGM/KM function, and (2) for every AGM/KM function there is a model (in the appropriate class) whose associated partial belief function coincides with it. The difference between revision and update boils down to two semantic properties that appear in a stronger form in revision and a weaker form in update.
\par
Section \ref{SEC:BeliefFunctions} reviews and compares the notions of KM belief update function and  AGM belief revision function. Section \ref{SEC:Frames} introduces the semantics, Section \ref{SEC:CorrespondenceUpdate} provides frame characterization results for the KM axioms and provides a characterization of belief update, while Section \ref{SEC:CorrespondenceRevision} does the same for revision. Section \ref{SEC:comparison} compares the two notions of update and revision in light of the results of the previous section. Section \ref{SEC:literature} reviews related literature and Section \ref{SEC:Conclusion} concludes.
\section{Belief change functions}
\label{SEC:BeliefFunctions}
In what follows we shall use the symbol $\circ$ for general belief change functions, the symbol $\diamond$ for belief update functions and the symbol $\ast$ for belief revision functions.\\
We consider a propositional logic based on a countable set \texttt{At} of atomic formulas. We denote by $\Phi_0$ the set of Boolean formulas constructed from  \texttt{At} as follows: $\texttt{At}\subset \Phi_0$ and if $\phi,\psi\in\Phi_0$ then $\neg\phi$ and $\phi\vee\psi$ belong to $\Phi_0$. Define  $\phi\rightarrow\psi$, $\phi\wedge\psi$, and $\phi\leftrightarrow\psi$ in terms of $\lnot$ and $\vee$ in the usual way.
\par
Given a subset $K$ of $\Phi_0 $, its deductive closure $Cn(K)\subseteq\Phi_0$ is defined as follows: $\psi \in Cn(K)$ if and only if there exist $\phi _1,...,\phi _n\in K$ \ (with $n\geq 0$) such that $(\phi _1\wedge ...\wedge \phi _n)\rightarrow \psi $ is a tautology. A set $K\subseteq \Phi_0 $ is \textit{consistent} if $ Cn(K)\neq \Phi_0 $; it is \textit{deductively closed} if $K=Cn(K)$. Given a set $K\subseteq \Phi_0$ and a formula $\phi\in\Phi_0$, the \emph{expansion} of $K$ by $\phi$, denoted by $K+\phi$, is defined as follows: $K+\phi=Cn\left(K\cup\{\phi\}\right)$.
\par
Let $K\subseteq\Phi_0$ be a \emph{consistent and deductively closed} set representing the agent's initial beliefs and let $\Psi \subseteq \Phi_0 $ be a set of formulas representing possible inputs for belief change (e.g. informational inputs). A \emph{belief change function} based on $\Psi$ and $K$ is a function $\circ:\Psi \rightarrow 2^{\Phi_0 }$ (where $2^{\Phi_0}$ denotes the set of subsets of $\Phi_0 $) that associates with every formula $\phi \in \Psi $ a set $ K\circ\phi \subseteq \Phi_0 $, interpreted as the change in $K$ prompted by the input $\phi$. We follow the common practice of writing $K\circ\phi$ instead of $\circ(\phi)$ which has the advantage of making it clear that the belief change function refers to a given, \emph{fixed}, $K$. If $\Psi \neq \Phi_0 $ then $\circ$ is called a \emph{partial} belief change function, while if $\Psi =\Phi_0 $ then $\circ$ is called a \emph{full-domain} belief change function.
\begin{definition}
\label{brf extension}
Let $\circ:\Psi \rightarrow 2^{\Phi_0}$ be a partial belief change function (thus $\Psi\subsetneq\Phi_0$) and $\circ ':\Phi_0 \rightarrow 2^{\Phi_0}$ a full-domain belief change function. We say that $\circ '$ is an \emph{extension} of $\circ$ if $\circ '$ coincides with $\circ $ on the domain of $\circ $, that is, if, for every $\phi \in \Psi$, $ K\circ '\phi= K\circ\phi $.
\end{definition}
\subsection{Belief update functions}
\label{SEC:UpdateFunctions}
We consider the notion of belief update introduced by Katsuno and Mendelzon in \cite{KatMen91} and compare it to the notion of belief revision introduced by Alchourr\'{o}n, G\"{a}rdenfors and Makinson in \cite{AGM85}. The formalism in the two theories is somewhat different. In \cite{KatMen91} a belief state is represented by a sentence in a finite propositional calculus and belief update is modeled as a function over sentences, while in \cite{AGM85} a belief state is represented (as we did above) as a set of sentences. Furthermore, while \cite{KatMen91} allows for the possibility of inconsistent initial beliefs, \cite{AGM85} take as starting point a consistent belief set. We follow closely the axiomatization of belief update proposed by \cite{Pep93,Pepetal96}, which makes update directly comparable to revision (note, however, that \cite{Pep93,Pepetal96} only cover the case of "strong" update, where axioms $(K\diamond6)$ and $(K\diamond7)$ are replaced by $(K\diamond9)$: see Definition \ref{DEF:strong_update function} below).
\par
\begin{definition}
\label{DEF:update}
A \emph{belief update function},  based on the  consistent and deductively closed set $K$ (representing the initial beliefs), is a full domain belief change function $\diamond:\Phi_0\to 2^{\Phi_0}$ that satisfies the following axioms: $\forall \phi,\psi\in\Phi_0$,
\begin{enumerate}
\item[]$(K\diamond 0)$\quad $K\diamond \phi=Cn(K\diamond \phi)$.
\item[]$(K\diamond 1)$\quad $\phi\in K\diamond \phi$.
\item[]$(K\diamond 2)$\quad If $\phi\in K$ then $K\diamond \phi=K$.
\item[]$(K\diamond 3)$\quad $K\diamond \phi=\Phi_0$ if and only if $\phi$ is a contradiction.
\item[]$(K\diamond 4)$\quad If $\phi\leftrightarrow\psi$ is a tautology then $K\diamond \phi=K\diamond \psi$.
\item[]$(K\diamond 5)$\quad $K\diamond (\phi\wedge\psi)\subseteq(K\diamond \phi)+\psi$.
\item[]$(K\diamond 6$)\quad If $\psi\in K\circ\phi$ and $\phi\in K\circ\psi$ then $K\circ\phi=K\circ\psi$ .
\item[]$(K\diamond 7$)\quad If $K$ is complete\footnote{
A belief set $K$ is \emph{complete} if, for every formula $\phi\in\Phi_0$, either $\phi\in K$ or $\lnot\phi\in K$; if $K$ is consistent, then $K$ is complete if and only if $\llbracket K\rrbracket$ is a singleton, where $\llbracket K\rrbracket$ denotes the set of maximally consistent sets of formulas (see Footnote \ref{FT:MCS}) that satisfy all the formulas in $K$.
}\,
then $(K\circ\phi)\cap(K\circ\psi)\subseteq K\circ(\phi\vee\psi)$.
    \end{enumerate}
\end{definition}
\begin{remark}
\label{REM:aboutKM8}
Katsuno and Mendelzon (\cite{KatMen91}) provide an additional axiom (they name it U8), which they call the "disjunction rule". Peppas et al. (\cite{Pepetal96}) translate it into the following "axiom", which makes use of maximally consistent sets of formulas (MCS).\footnote{\label{FT:MCS}
A set of formulas $\Delta$ is maximally consistent if it is consistent and, furthermore, if $\Gamma$ is consistent and $\Delta\subseteq\Gamma$ then $\Delta=\Gamma$. Every MCS is deductively closed and complete.
}\ 
 Given a set of formulas $\Gamma$, let $\llbracket\Gamma\rrbracket$ be the set of MCS that contain all the formulas in $\Gamma$. The additional axiom is the following:
\begin{equation}
\tag{$K\diamond 8$}
\label{KM8}
  \text{If } \llbracket K\rrbracket\ne\varnothing  \text{ then } K\diamond\phi=\bigcap\limits_{w\in \llbracket K\rrbracket} w\diamond\phi.
\end{equation}
($K\diamond 8$) is of a different nature than the other axioms, since it applies the update operator not only to the initial belief set $K$ but also to the individual MCS contained in $ \llbracket K\rrbracket$. It seems that ($K\diamond 8$) is more of a condition on the interpretation or semantics than a real axiom; it is not needed in our framework since its role is directly captured by the semantics detailed in the next section. We shall return to ($K\diamond 8$) in Section \ref{SEC:CorrespondenceUpdate}.
\end{remark}
\noindent
$(K\diamond0)$ does not appear in the list of axioms provided by Katsuno and Mendelzon, since their formalism is not in terms of belief sets. For $i\in\{1,2,4,5,6\}$, axiom $(K\diamond i)$ is a translation of Katsuno and Mendelzon's axiom $(Ui)$ (for details see \cite{Pep93}). $(K\diamond 3)$ is the translation of Katsuno and Mendelzon's axiom $(U3)$ when attention is restricted to the case where the initial belief set $K$ is consistent.\footnote{ 
Katsuno and Mendelzon allow for the possibility that the initial beliefs are inconsistent in which case ($K\diamond 3$) would be stated as follows: $K\diamond \phi=\Phi_0$ if and only if either $K$ is inconsistent or $\phi$ is inconsistent. In order to facilitate the comparison between belief update and belief revision, we follow \cite{Pepetal96} and restrict attention to the case where the initial belief set $K$ is consistent. It should be noted, however, that one important difference between update and revision is precisely that updating an inconsistent $K$ by a consistent formula $\phi$ yields the inconsistent belief set $\Phi_0$, while revising an inconsistent $K$ by a consistent formula $\phi$ yields a consistent set (AGM axiom $K\ast5$).
} 
The following Lemma will be used later.
\begin{lemma}
\label{LEM:*3forKM}
Assuming that $K$ is deductively closed $(K=Cn(K))$, every belief update function satisfies the following axiom: $$K\diamond\phi\subseteq K+\phi.$$
\end{lemma}
{\small
\begin{proof}
Since $((\phi\vee\lnot\phi)\wedge\phi)\leftrightarrow\phi$ is a tautology, by $(K\diamond4)$,
\begin{equation}
\label{EQ:*3forKM_1}
K\diamond((\phi\vee\lnot\phi)\wedge\phi)=K\diamond\phi.
\end{equation}
By $(K\diamond5)$,
\begin{equation}
\label{EQ:*3forKM_2}
K\diamond((\phi\vee\lnot\phi)\wedge\phi)\subseteq(K\diamond(\phi\vee\lnot\phi))+\phi.
\end{equation}
Thus, by \eqref{EQ:*3forKM_1} and  \eqref{EQ:*3forKM_2},
\begin{equation}
\label{EQ:*3forKM_3}
K\diamond\phi\subseteq(K\diamond(\phi\vee\lnot\phi))+\phi.
\end{equation}
Since, by hypothesis, $K$ is deductively closed and $(\phi\vee\lnot\phi)$ is a tautology,  $(\phi\vee\lnot\phi)\in K$, so that, by $(K\diamond2)$, $K\diamond(\phi\vee\lnot\phi)=K$. It follows from this and \eqref{EQ:*3forKM_3} that $K\diamond\phi\subseteq K+\phi.$
\end{proof}
}
\begin{remark}
\label{REM:KM9}
KM show that their notion of belief update corresponds, semantically, to partial pre-orders on the set of possible worlds (maximally consistent sets of formulas); furthermore, they show that if their axioms $(U6)$ and $(U7)$ are replaced by a stronger axiom, which they call $(U9)$, then belief update corresponds semantically to total pre-orders on the set of possible worlds. The translation of their axiom $(U9)$ in our framework is the following axiom (see \cite{Pepetal96}):
\begin{equation}
\tag{$K\diamond 9$}
\label{KM9}
  \text{If } K \text{ is complete and }\lnot\psi\notin K\diamond\phi \text{ then }(K\diamond\phi)+\psi\subseteq K\diamond (\phi\wedge\psi).
  \end{equation}
\end{remark}
\begin{definition}
\label{DEF:strong_update function}
A \emph{strong belief update function} is a full-domain belief change function $\diamond:\Phi_0\to 2^{\Phi_0}$ that satisfies axioms $(K\diamond0), (K\diamond1), (K\diamond2), (K\diamond3), (K\diamond4), (K\diamond5), (K\diamond9)$.
\end{definition}
\subsection{Belief revision functions}
\label{SEC:RevisionFunctions}
In this section we consider the notion of belief revision proposed Alchourr\'{o}n, G\"{a}rdenfors and Makinson in \cite{AGM85}.
\begin{definition}
\label{DEF:revision}
A \emph{belief revision function}, based on the  consistent and deductively closed set $K$ (representing the initial beliefs), is a full domain belief change function $\ast:\Phi_0\to 2^{\Phi_0}$ that satisfies the following axioms: $\forall \phi,\psi\in\Phi_0$,
\begin{enumerate}
\item[]($K\ast 1$)\quad $K\ast\phi=Cn(K\ast\phi)$.
\item[]($K\ast 2$)\quad $\phi \in K\ast\phi$.
\item[]($K\ast 3$)\quad $K\ast\phi\subseteq K+ \phi$.
\item[]($K\ast 4$) if $\lnot \phi \notin K$, then $K\subseteq K\ast\phi$.
\item[]($K\ast 5$) $K\ast\phi=\Phi_0$ if and only if $\phi $ is a contradiction.
\item[]($K\ast 6$) if $\phi \leftrightarrow \psi $ is a tautology then $K\ast\phi=K\ast\psi$.
\item[]($K\ast 7$) $K\ast(\phi \wedge \psi)\subseteq (K\ast\phi)+\psi$.
\item[]($K\ast 8$) if $\lnot \psi \notin K\ast\phi$, then $(K\ast\phi)+\psi\subseteq K\ast(\phi \wedge \psi).$
\end{enumerate}
\end{definition}\bigskip
\begin{remark}
\label{REMcompare-circ-diamond}
Note that
\begin{itemize}
\item $(K\ast 1)$ coincides with $(K\diamond 0)$,
\item $(K\ast 2)$ coincides with $(K\diamond 1)$,
\item $(K\ast 3)$ applies also to belief update (Lemma \ref{LEM:*3forKM}),
\item $(K\ast 5)$ coincides with $(K\diamond 3)$,
\item $(K\ast 6)$ coincides with $(K\diamond 4)$,
\item $(K\ast 7)$ coincides with $(K\diamond 5)$
\end{itemize}
\noindent
On the other hand,
\begin{itemize}[label=$\triangleright$]
\item $(K\ast 4)$ is a stronger version of $(K\diamond 2)$, and
\item $(K\ast 8)$ is a stronger version of $(K\diamond 9)$.
\end{itemize}
Thus one can view the AGM theory of belief revision as a stronger version of the strong version of KM belief update, namely that which satisfies axioms $(K\diamond0)$-$(K\diamond5)$ and $(K\diamond9)$. At the semantic level, this point is discussed in detail in Section \ref{SEC:CorrespondenceRevision}.
\end{remark}
\section{Semantics: Kripke-Lewis frames}
\label{SEC:Frames}
In this section we introduce semantic structures and establish a correspondence between axioms of belief change functions and properties of the structures.
\par
\begin{definition}
\label{DEF:frame}
A \emph{Kripke-Lewis frame} is a triple $\left\langle {S,\mathcal B,f} \right\rangle$ where
\begin{enumerate}
\item $S$ is a set of \emph{states}; subsets of $S$ are called  \emph{events}.
\item $\mathcal B \subseteq S \times S$ is a binary \emph{belief relation} on $S$ which is serial: $\forall s\in S, \exists s'\in S$, such that $s\mathcal B s'$. We denote by $\mathcal B(s)$ the set of states that are reachable from $s$ by $\mathcal B$: $\mathcal B(s)=\{s'\in S: s\mathcal B s'\}$. $\mathcal B(s)$ is interpreted as the set of states that the agent considers doxastically possible at state $s$.
\item $f:S\times (2^S\setminus\varnothing) \rightarrow 2^S$ is a \emph{selection function} that associates with every state-event pair $(s,E)$ (with $E\ne\varnothing$) a set of states $f(s,E)\subseteq S$, such that
    \begin{enumerate}
    \item[] (3.1)\quad $ f(s,E)\ne\varnothing$ (Consistency).
    \item[] (3.2)\quad $ f(s,E)\subseteq E$ (Success).
    \item[] (3.3)\quad if $s\in E$ then $s\in f(s,E)$ (Weak Centering).
    \end{enumerate}
    $ f(s,E)$ is interpreted as the set of $E$-states that are closest (or most similar) to $s$ (an $E$-state is a state that belongs to $E$).\\
    (3.1) says that there is at least one $E$-state that is closest to $s$ (note that, by hypothesis, $E\ne\varnothing$).\\
    (3.2) is a coherence requirement: the states that are closest to $s$, conditional on $E$, are indeed $E$-states.\\
    (3.3) says that if $s$ is an $E$-state then it is one of the closest $E$-states to itself.\footnote{
     A stronger property, called \emph{centering}, requires that if $s\in E$ then $f(s,E)=\{s\}$. For our purposes this stronger property is not needed.
     }
    \end{enumerate}
\end{definition}
Adding a valuation to a frame yields a model. Thus a \emph{model} is a tuple $\left\langle {S,\mathcal B,f,V} \right\rangle$ where $\left\langle {S,\mathcal B,f} \right\rangle$ is a frame and $V:\texttt{At}\rightarrow 2^S$ is a valuation that assigns to every atomic formula $p\in\texttt{At}$ the set of states where $p$ is true.
Given a model $M=\left\langle {S,\mathcal B,f,V} \right\rangle$ define truth of a Boolean formula $\phi\in\Phi_0$ at a state $s\in S$ in model $M$, denoted by $s\models_M\phi$, in the usual way.
\begin{definition}
\label{Truth0}
Truth of a formula at a state is defined as follows:
\begin{enumerate}
\item if $p\in\texttt{At}$ then $s\models_M p$ if and only if $s\in V(p)$,
\item $s\models_M\neg\phi$ if and only if $s\not\models_M\phi$,
\item $s\models_M(\phi\vee\psi)$ if and only if $s\models_M\phi$ or $s\models_M\psi$ (or both).
\end{enumerate}
\end{definition}
\noindent
We denote by $\Vert\phi\Vert_M$ the truth set of formula $\phi$ in model $M$: $\Vert\phi\Vert_M=\{s\in S: s\models_M\phi\}$.
\par
Given a model $M=\left\langle {S,\mathcal B,f,V} \right\rangle$ and a state $s\in S$,  let $K_{s,M}=\{\phi \in \Phi _{0}:\mathcal B(s) \subseteq \Vert \phi \Vert_{M} \}$; thus a Boolean formula $\phi$ belongs to $K_{s,M}$ if and only if at state $s$ the agent believes $\phi$ (in the sense that $\phi$ is true at every state that the agent considers doxastically possible at state $s$). We identify $K_{s,M}$ with the agent's \emph{initial beliefs at state} $s$. It is shown in Lemma \ref{LEM:KfromModel} below that the set $K_{s,M}\subseteq\Phi_0$ so defined is deductively closed and consistent.
\par
It is possible that, in a given model $M$,  $s'\in\mathcal B(s)$ and $K_{s,M}\ne K_{s',M}$, that is, it is possible that the agent's initial beliefs at state $s$ are different from the agent's beliefs at a state $s'$ that is doxastically accessible from $s$. Such a phenomenon can be ruled out by imposing two additional properties on $\mathcal B$: transitivity (if $s'\in\mathcal B(s)$ then $\mathcal B(s')\subseteq \mathcal B(s)$) and euclideanness (if $s'\in\mathcal B(s)$ then $\mathcal B(s)\subseteq \mathcal B(s')$). Since none of the results proved below require these two additional properties of $\mathcal B$, we did not incorporate them in Definition \ref{DEF:frame}.  Note also that we did \emph{not} assume reflexivity of $\mathcal B$.\footnote{
$\mathcal B$ is reflexive if, $\forall s\in S, s\in\mathcal B(s)$.
}\ 
Thus, we allow for the possibility of false beliefs (that is, it is possible in a model to have, at some state $s$ and for some formula $\phi$, $\mathcal B(s)\subseteq\Vert\phi\Vert_M$ and also $s\notin\Vert\phi\Vert_M$). \smallskip
\par
 Next, given a model $M=\left\langle {S,\mathcal B,f,V} \right\rangle$ and a state $s\in S$, let $\Psi_M=\{\phi\in\Phi_0:\Vert\phi\Vert_M\ne\varnothing\}$\footnote{
Since, in any given model there are formulas $\phi$ such that $\Vert\phi\Vert_M=\varnothing$ (at the very least all the contradictions), $\Psi_M$ is a proper subset of $\Phi_0$.
  }\ 
and define the following partial belief change function
 $\circ:\Psi_M\to 2^{\Phi_0}$, based on $K_{s,M}$ and $\Psi_M$:
\begin{equation}
\tag{RI}
\label{RI}
\begin{array}{*{20}{l}}
 \psi\in K_{s,M}\circ \phi\,\,\text{ if and only if }\,\, \forall s'\in\mathcal B(s), f\left(s',\Vert\phi\Vert_M\right)\subseteq\Vert\psi\Vert_M.
\end{array}
\end{equation}
Given the customary interpretation of selection functions in terms of conditionals, (RI) can be interpreted as stating that $\psi\in K_{s,M}\circ\phi$ if and only if at state $s$ the agent believes that "if $\phi$ is (were) the case then $\psi$ is (would be) the case". `RI' stand for `Ramsey Interpretation'.\footnote{\label{FT:Ramsey}
In the literature the expression `Ramsey Test' is used to refer to the following passage from \cite[p. 247]{Ram50}: "If two people are arguing "If $p$ will $q$?" and are both in doubt as  to $p$, they are adding $p$ hypothetically to their stock of knowledge and arguing on that  basis about $q$".
}\ 
\begin{lemma}
\label{LEM:KfromModel}
Let $F=\left\langle {S,\mathcal B,f} \right\rangle$ be a frame, $M$ a model based on $F$, $s\in S$ a state, and $\Psi_M=\{\phi\in\Phi_0:\Vert\phi\Vert_M\ne\varnothing\}$. Let $K_{s,M}=\{\psi \in \Phi _{0}:\mathcal B(s) \subseteq \Vert \psi \Vert_M \}$ be the initial beliefs at state $s$ and, for every $\phi\in\Psi_M$, let $K_{s,M}\circ\phi\subseteq\Phi_0$ be the new beliefs (in response to input $\phi$) defined by \eqref{RI}. Then
\begin{enumerate}[label=(\Alph*)]
\item the set $K_{s,M}$ is consistent and deductively closed, and
\item the set \,$K_{s,M}\circ\phi$ \,is consistent and deductively closed.
\end{enumerate}
\end{lemma}
{\small
\begin{proof}
To simplify the notation, we omit the subscript $M$ referring to the given model, that is, we write $K_s$ instead of $K_{s,M}$ and $\Vert\phi\Vert$ instead of $\Vert\phi\Vert_M$.
\begin{enumerate}[label=(\Alph*)]
\item First we show that $K_s$ is deductively closed, that is, $K_s=Cn(K_s)$. If $\psi \in K_s$ then $\psi \in Cn(K_s)$, because $\psi \rightarrow \psi $ is a tautology; thus $K_s\subseteq Cn(K_s)$. To show that $Cn(K_s)\subseteq K_s$, let $\psi \in Cn(K_s)$, that is, there exist $\phi _{1},...,\phi _{n}\in K_s$ ($n\geq 0$) such that $( \phi _{1}\wedge ...\wedge \phi _{n}) \rightarrow \psi $ is a tautology. Since $\Vert \phi _{1}\wedge ...\wedge \phi _{n}\Vert =\Vert \phi _{1}\Vert \cap ...\cap \Vert \phi _{n}\Vert $ and,  for all $i=1,...,n$, $\phi _{i}\in K_s$ (that is, $\mathcal B(s)\subseteq \Vert \phi _{i}\Vert $), it follows that $\mathcal B(s) \subseteq \Vert \phi _{1}\wedge ...\wedge \phi _{n}\Vert $. Since $( \phi _{1}\wedge ...\wedge \phi _{n}) \rightarrow \psi $ is a tautology, $\Vert ( \phi _{1}\wedge ...\wedge \phi _{n}) \rightarrow \psi \Vert =S $, that is, $\Vert \phi _{1}\wedge ...\wedge \phi _{n}\Vert \subseteq \Vert \psi \Vert .$ Thus $\mathcal B(s)\subseteq \Vert \psi \Vert $, that is, $\psi \in K_s$. \\
    Next we show that $K_s$ is consistent, that is, $Cn(K_s)\neq \Phi _{0}$. Let $p\in\texttt{At}$ be an atomic formula. Then $\Vert p\wedge \lnot p\Vert =\varnothing $. By seriality of $\mathcal B$, $\mathcal B(s)\neq\varnothing$ so that $\mathcal B(s)\nsubseteq \Vert p\wedge \lnot p\Vert $, that is, $(p\wedge \lnot p)\notin K_s$ and hence, since $K_s=Cn(K_s)$, $(p\wedge \lnot p)\notin Cn(K_s)$.
    \item First we show that $K_s\circ\phi$ is deductively closed, that is, $K_s\circ\phi= Cn\left(K_s\circ\phi\right)$. The inclusion $K_s\circ\phi\subseteq Cn\left(K_s\circ\phi\right)$ follows from the fact that, for every $\psi\in K_s\circ\phi$, $\psi\rightarrow\psi$ is a tautology. Next we show that $Cn\left(K_s\circ\phi\right)\subseteq K_s\circ\phi$.  Since, by hypothesis, $\Vert \phi\Vert \ne \varnothing$,  $f(s',\Vert\phi\Vert)$ is defined for every $s'\in \mathcal B(s)$. Fix an arbitrary $\psi \in Cn\left(K_s\circ\phi\right)$; then there exist $\phi _{1},...,\phi _{n}\in K_s\circ\phi$ ($n\geq 0$) such that $( \phi _{1}\wedge ...\wedge \phi _{n}) \rightarrow \psi $ is a tautology, so that $\Vert( \phi _{1}\wedge ...\wedge \phi _{n}) \rightarrow \psi  \Vert = S$, that is, $\Vert\phi _{1}\wedge ...\wedge \phi _{n}\Vert\subseteq\Vert\psi\Vert$. Fix an arbitrary $s'\in\mathcal B(s)$ and an arbitrary $i=1,...,n$. Then, since $\phi _{i}\in K_s\circ\phi$, by \eqref{RI} $f(s',\Vert\phi\Vert)\subseteq\Vert\phi_i\Vert$. Hence $f(s',\Vert\phi\Vert)\subseteq\Vert\phi _{1}\Vert\cap...\cap\Vert\phi _{n}\Vert=\Vert\phi _{1}\wedge ...\wedge \phi _{n}\Vert$. Since $\Vert\phi _{1}\wedge ...\wedge \phi _{n}\Vert\subseteq\Vert\psi\Vert$ it follows that $f(s',\Vert\phi\Vert)\subseteq\Vert\psi\Vert$. Thus, since $s'\in\mathcal B(s)$ was chosen arbitrarily,  $\psi\in K_s\circ\phi$.\\
        Next we show that $K_s\circ\phi$ is consistent, that is, $Cn(K_s\circ\phi)\neq \Phi _{0}$. Let $p\in\texttt{At}$ be an atomic formula. Then $\Vert p\wedge \lnot p\Vert =\varnothing $. Since, by hypothesis, $\Vert \phi\Vert \ne \varnothing$, by Property (3.1) of the definition of frame (Definition \ref{DEF:frame}) $f(s',\Vert\phi\Vert)\ne\varnothing$, for every $s'\in\mathcal B(s)$ (recall that, by seriality of $\mathcal B$, $\mathcal B(s)\ne\varnothing$). Thus $f(s',\Vert\phi\Vert)\nsubseteq\Vert p\wedge \lnot p\Vert$ so that $(p\wedge\lnot p)\notin K_s\circ\phi$. Hence, since (as shown above) $K_s\circ\phi=Cn\left(K_s\circ\phi\right)$, $(p\wedge\lnot p)\notin Cn\left(K_s\circ\phi\right)$ so that $Cn\left(K_s\circ\phi\right)\ne\Phi_0$.
\end{enumerate}
\end{proof}
}
In what follows, when stating an axiom for a belief change function, we implicitly assume that it applies to every formula \emph{in its domain}. For example, the axiom $\phi\in K\circ\phi$ asserts that, for all $\phi$ in the domain of $\circ$, $\phi\in K\circ\phi$.
\begin{definition}
\label{DEF:}
An axiom is \emph{valid on a frame $F$} if, for every model based on that frame and for every state $s$ in that model, the partial belief change function defined by \eqref{RI} satisfies the axiom. An axiom is \emph{valid on a set of frames $\mathcal F$} if it is valid on every frame $F\in\mathcal F$.
\end{definition}
\begin{proposition}
\label{PROP:validaxioms}
The following axioms are valid on the set of all frames (as defined in Definition \ref{DEF:frame}):
\begin{align}
  1. &\quad K\circ\phi\subseteq K+\phi.\tag{$\ast3$}\\
  2. & \quad\phi\in K\circ\phi. \tag{$\diamond1/\ast2$}\\
  3. & \quad\text{if } \phi\leftrightarrow\psi \text{ is a tautology, then } K\circ\phi = K\circ\psi.\tag{$\diamond4/\ast6$}
  \end{align}
\end{proposition}
{\small
\begin{proof}
 Let $F$ be a frame, $M$ a model based on $F$ and $s\in S$ a state in that model. Let $\Psi_M=\{\phi\in\Phi_0:\Vert\phi\Vert_M\ne\varnothing\}$, $K_{s,M}=\{\phi\in\Phi_0:\mathcal B(s)\subseteq\Vert\phi\Vert_M \}$ and, for every $\phi\in\Psi_M$, let $K_{s,M}\circ\phi$ be the partial belief change function defined by \eqref{RI}.  As before, in what follows we simplify the notation by omitting the subscript $M$ referring to the given model.
\begin{enumerate}
 \item Let $\phi\in\Psi_0$ be in the domain of $\circ$ (that is, $\Vert\phi\Vert\ne\varnothing$) and fix an arbitrary $\psi\in K_s\circ\phi$. Then, by \eqref{RI}, $\forall s'\in\mathcal B(s)$, $f(s',\Vert\phi\Vert)\subseteq \Vert\psi\Vert$. Thus, since $\Vert\psi\Vert\subseteq \Vert\lnot\phi\Vert\cup\Vert\psi\Vert=\Vert\phi\rightarrow\psi\Vert$ we have that
      \begin{equation}
      \label{zero}
      \forall s'\in\mathcal B(s),\,f(s',\Vert\phi\Vert)\subseteq\Vert\phi\rightarrow\psi\Vert.
      \end{equation}
 We need to show that $\psi\in K_s+\phi=Cn(K_s\cup\{\phi\})$, that is -- since, by (A) of Lemma \ref{LEM:KfromModel}, $K_s=Cn(K_s)$ -- that $(\phi\rightarrow\psi)\in K_s$, i.e. $\mathcal B(s)\subseteq\Vert\phi\rightarrow\psi\Vert$. Fix an arbitrary $s'\in\mathcal B(s)$. If $s'\notin\Vert\phi\Vert$ then $s'\in\Vert\lnot\phi\Vert\subseteq\Vert\lnot\phi\Vert\cup\Vert\psi\Vert=
 \Vert\phi\rightarrow\psi\Vert$. If $s'\in\Vert\phi\Vert$, then by Property (3.2) of the definition of frame (Definition \ref{DEF:frame}), $s'\in f(s',\Vert\phi\Vert)$, so that, by \eqref{zero}, $s'\in\Vert\phi\rightarrow\psi\Vert$.
  \item Fix an arbitrary $\phi\in\Phi_0$ such that $\Vert\phi\Vert\ne\varnothing$ and an arbitrary $s'\in\mathcal B(s)$. By Property (3.2) of the definition of frame (Definition \ref{DEF:frame}), $f(s',\Vert\phi\Vert)\subseteq\Vert\phi\Vert$. Hence, by \eqref{RI}, $\phi\in K\circ\phi$.
  \item Let $\phi,\psi\in\Phi_0$ be such that $\Vert\phi\Vert\ne\varnothing$ and $\Vert\psi\Vert\ne\varnothing$. If $\phi\leftrightarrow\psi$ is a tautology then $\Vert\phi\leftrightarrow\psi\Vert=S$, that is, $\Vert\phi\Vert=\Vert\psi\Vert$, so that, for every $s'\in\mathcal B(s)$, $f(s',\Vert\phi\Vert)=f(s',\Vert\psi\Vert)$. Hence, for every $\chi\in\Psi$, $f(s',\Vert\phi\Vert)\subseteq\Vert\chi\Vert$ if and only if $f(s',\Vert\psi\Vert)\subseteq\Vert\chi\Vert$, that is, by \eqref{RI}, $\chi\in K\circ\phi$ if and only if $\chi\in K\circ\psi$.
  \end{enumerate}
\end{proof}
}
\section{Frame correspondence. Part 1: Update}
\label{SEC:CorrespondenceUpdate}
A stronger notion than validity is that of frame correspondence. The following definition mimics the notion of frame correspondence in modal logic.
\begin{definition}
\label{DEF:}
We say that an axiom $A$ of belief change functions is \emph{characterized by} or \emph{corresponds to} a property $P$ of frames if the following is true:
\begin{enumerate}[label=(\arabic*)]
\item axiom $A$ is valid on the class of frames that satisfy property $P$, and
\item if a frame does not satisfy property $P$ then axiom $A$ is not valid on that frame, that is, there is a model based on that frame and a state in that model where the partial belief change function defined by \eqref{RI} violates axiom $A$.
    \end{enumerate}
\end{definition}

\begin{proposition}
\label{PROP:K_circ_2}
The following axiom:
\begin{equation}
\tag{$\diamond2$}
\label{KM2}
\text{if } \phi\in K\text{ then }K\circ\phi=K
\end{equation}
is characterized by the following property of frames: $\forall s\in S,\forall E\subseteq S$,
\begin{equation}
\tag{$P_{\diamond2}$}
\label{PKM2}
 \text{ if } \mathcal B(s)\subseteq E\text{ then, }\forall s'\in\mathcal B(s), f(s',E)\subseteq\mathcal B(s).
\end{equation}
\end{proposition}
{\small
\begin{proof}
(A) Fix a frame that satisfies property \eqref{PKM2}, an arbitrary model $M$ based on it and an arbitrary state $s\in S$. As before, we simplify the notation and omit the subscript $M$ referring to the given model. Let $\phi\in\Phi_0$ be such that $\phi\in K_s$, that is, $\mathcal B(s)\subseteq\Vert \phi\Vert$ (note that, by seriality of $\mathcal B$, it follows that $\Vert \phi\Vert \ne \varnothing$  so that $f(s',\Vert\phi\Vert)$ is defined for every $s'$).\\
First we show that $K_s\subseteq K_s\circ\phi$. Let $\psi\in K_s$. Then $\mathcal B(s)\subseteq \Vert\psi\Vert$. By property  \eqref{PKM2} (with $E=\Vert\phi\Vert$), for every $s'\in\mathcal B(s)$, $f(s',\Vert\phi\Vert)\subseteq\mathcal B(s)$ and thus $f(s',\Vert\phi\Vert)\subseteq\Vert\psi\Vert$, so that, by \eqref{RI}, $\psi\in K_s\circ\phi$. \\
Conversely, suppose that $\psi\in K_s\circ\phi$, that is, $\forall s'\in\mathcal B(s),\,f(s',\Vert\phi\Vert)\subseteq\Vert\psi\Vert$, so that  $\bigcup\limits_{s'\in\mathcal B(s)} { f(s',\Vert \phi\Vert)}\subseteq\Vert\psi\Vert$. Since $\mathcal B(s)\subseteq\Vert \phi\Vert$, for every $s'\in \mathcal B(s)$, $s'\in\Vert \phi\Vert$ and thus, by Property (3.3) of Definition \ref{DEF:frame} (Weak Centering), $\{s'\}\subseteq f(s',\Vert \phi\Vert)$. Hence $\mathcal B(s)=\bigcup\limits_{s'\in\mathcal B(s)} \{s'\} \subseteq \bigcup\limits_{s'\in\mathcal B(s)} { f(s',\Vert \phi\Vert)}\subseteq\Vert\psi\Vert$; it follows from this and the definition of $K_s$ that $\psi\in K_s$. \smallskip\\
(B) Fix a frame that violates property \eqref{PKM2}. Then there exist two states $s,s'\in S$ and an event $E\subseteq S$ such that (a) $\mathcal B(s)\subseteq E$, (b) $s'\in\mathcal B(s)$ and (c) $f(s',E)\not\subseteq\mathcal B(s)$. Let $p,q\in\texttt{At}$ be atomic formulas and construct a model where $\Vert p\Vert=E$ and $\Vert q\Vert=\mathcal B(s)$. Since $\mathcal B(s)\subseteq \Vert q\Vert$,  $q\in K_s$.  Since  $f(s',E)\not\subseteq\mathcal B(s)$, $f(s',\Vert p\Vert)\not\subseteq\Vert q\Vert$ and thus, since $s'\in\mathcal B(s)$, by \eqref{RI} $q\notin K_s\circ p\,$ so that $K_s\circ p\ne K_s$.
\end{proof}
}
\begin{proposition}
\label{PROP:KM5/AGM7}
The following axiom:
\begin{equation}
\tag{$\diamond5/\ast7$}
\label{KM5/AGM7}
K\circ(\phi\wedge\psi) \subseteq (K\circ\phi)+\psi
\end{equation}
is characterized by the following property of frames: $\forall s\in S,\forall E,F,G\in 2^S$, with $E\cap F\ne\varnothing$,
\begin{equation}
\tag{$P_{\diamond5/\ast7}$}
\label{PKM5AGM7}
\renewcommand{\arraystretch}{1.3}
\begin{array}{l}
\text{if, } \forall s'\in \mathcal B(s), \,f(s',E\cap F)\subseteq G,\\
\text{then, } \forall s'\in \mathcal B(s),  \,f(s',E)\cap F\subseteq G.
\end{array}
\end{equation}
\end{proposition}
{\small
\begin{proof}
(A) Fix a frame that satisfies property \eqref{PKM5AGM7}, an arbitrary model $M$ based on it and an arbitrary state $s\in S$ and let $\circ$ be the belief change function defined by \eqref{RI}.  Let $\phi,\psi\in\Phi_0$ be such that $\phi\wedge\psi$ is in the domain of $\circ$, that is, $\Vert\phi\wedge\psi\Vert\ne\varnothing$ (so that, since $\Vert\phi\wedge\psi\Vert=\Vert\phi\Vert\cap\Vert\psi\Vert$, also $\phi$ and $\psi$ are in the domain of $\circ$). We want to show that $K_s\circ(\phi\wedge\psi)\subseteq (K_s\circ\phi) +\psi$ (recall that $(K_s\circ\phi) +\psi=Cn\left((K_s\circ\phi)\cup\{\psi\} \right)$. By (A) of Lemma \ref{LEM:KfromModel}, $K_s\circ\phi$ is deductively closed, so that, $\forall\chi\in\Phi_0$,
\begin{equation}
\label{uno}
\setlength{\arraycolsep}{2pt}
\begin{array}{lll}
\chi\in Cn\left((K_s\circ\phi)\cup\{\psi\} \right)&\quad\text{if and only if}&(\psi\rightarrow\chi)\in K_s\circ\phi\\
\text{if and only if},\,\forall s'\in\mathcal B(s),&f(s',\Vert\phi\Vert)\subseteq\Vert\psi\rightarrow\chi\Vert&=\Vert\lnot\psi\Vert\cup\Vert\chi\Vert\\
&&=\left(S\setminus\Vert\psi\Vert\right)\cup\Vert\chi\Vert.
\end{array}
\end{equation}
First note that
\begin{equation}
\label{EQ:equuiv}
f(s',\Vert\phi\Vert)\subseteq(S\setminus\Vert\psi\Vert)\cup \Vert\chi\Vert \text{ if and only if } f(s',\Vert\phi\Vert)\cap\Vert\psi\Vert\subseteq\Vert\chi\Vert.
\end{equation}
Fix an arbitrary $\chi\in K_s\circ(\phi\wedge\psi)$. Then, by \eqref{RI}, $\forall s'\in\mathcal B(s)$, $f(s',\Vert\phi\Vert\cap\Vert\psi\Vert)\subseteq\Vert\chi\Vert$. By Property  \eqref{PKM5AGM7} and by \eqref{EQ:equuiv}, $\forall s'\in\mathcal B(s),\, f(s',\Vert\phi\Vert)\subseteq (S\setminus \Vert\psi\Vert)\cup \Vert\chi\Vert=\Vert \psi\rightarrow \chi\Vert$ and thus, by \eqref{uno}, $\chi\in Cn\left((K_s\circ\phi)\cup\{\psi\} \right)$.\\[3pt]
(B) Fix a frame that violates property \eqref{PKM5AGM7}. Then there exist two states $s,\hat s\in S$ and three events $E, F$ and $G$, with $E\cap F\ne\varnothing$, such that (a) $\forall s'\in\mathcal B(s),\,f(s',E\cap F)\subseteq G$, (b) $\hat s\in\mathcal B(s)$ and (c) $f(\hat s,E)\cap F\nsubseteq G$, that is, by \eqref{EQ:equuiv}, $f(\hat s,E) \nsubseteq (S\setminus F)\cup G$. Let $p,q,r\in\texttt{At}$ be atomic formulas and construct a model where $\Vert p\Vert=E$,  $\Vert q\Vert=F$ and $\Vert r\Vert=G$. Then, by (a), $r\in K_s\circ(p\wedge q)$ and, by (c), $f(\hat s,\Vert p \Vert)\nsubseteq \Vert q\rightarrow r\Vert$, so that, by (b) and \eqref{RI}, $(q\rightarrow r)\notin K_s\circ p$ and thus, by \eqref{uno}, $r\notin Cn\left((K_s\circ p)\cup\{q\} \right)$.
\end{proof}
}
\begin{remark}
\label{REM:P'5/7}
A stronger property than \eqref{PKM5AGM7} is the following:
\begin{equation}
\tag{$P'_{\diamond5/\ast7}$}
\label{P'5/7}
\renewcommand{\arraystretch}{1.3}
\begin{array}{c}
\forall s\in S,\forall E,F\in 2^S, \forall s'\in \mathcal B(s),\\
f(s',E)\cap F\subseteq f(s',E\cap F).
\end{array}
\end{equation}
\end{remark}
\noindent
To see that \eqref{P'5/7} implies \eqref{PKM5AGM7}, let $G\subseteq S$ be such that $\forall s'\in \mathcal B(s)$, $f(s',E\cap F)\subseteq G$. By \eqref{P'5/7}, $\forall s'\in \mathcal B(s)$, $f(s',E)\cap F\subseteq f(s',E\cap F)$, so that $f(s',E)\cap F\subseteq G$. On the other hand, \eqref{PKM5AGM7} does not imply \eqref{P'5/7}, as the following example shows: $S=\{s_0,s_1,s_2,s_3,s_4,s_5\},\,\mathcal B(s_0)=\{s_1,s_2\},\,E=\{s_3,s_4,s_5\},\,F=f(s_1,E)=\{s_3,s_4\},\, f(s_1,E\cap F)=\{s_3\},\,f(s_2,E)=f(s_2,E\cap F)=\{s_4\}$. Then $f(s_1,E\cap F)\cup f(s_2,E\cap F)=\{s_3,s_4\}$ so that any $G\subseteq S$ that contains $f(s_1,E\cap F)\cup f(s_2,E\cap F)$ must be a superset of $\{s_3,s_4\}$ and we do have that both $f(s_1,E)\cap F\subseteq G$ and $f(s_2,E)\cap F\subseteq G$ are satisfied, while $\{s_3,s_4\}=f(s_1,E)\cap F\nsubseteq f(s_1,E\cap F)=\{s_3\}$.\smallskip
\par\smallskip
From Proposition \ref{PROP:KM5/AGM7} and Remark \ref{REM:P'5/7} we get the following corollary.\smallskip
\begin{corollary}
Axiom \eqref{KM5/AGM7} is valid on the set of frames that satisfy Property \eqref{P'5/7}.
\end{corollary}\smallskip
\begin{proposition}
\label{PROP:KM6}
The following axiom:
\begin{equation}
\tag{$\diamond6$}
\label{EQ:KM6}
\text{if } \psi\in K\circ\phi \text{ and } \phi\in K\circ\psi \text{ then } K\circ\phi=K\circ\psi
\end{equation}
is characterized by the following property of frames: $\forall s\in S,\forall E,F\in 2^S\setminus\varnothing$,
\begin{equation}
\tag{$P_{\diamond6}$}
\label{EQ:PKM6}
\begin{array}{l}
 \text{ if, } \forall s'\in\mathcal B(s),\,  f(s',E)\subseteq F \text{ and }\, f(s',F)\subseteq E \\[5pt]
 \text{ then } \bigcup\limits_{s'\in\mathcal B(s)} {f(s',E)} =\bigcup\limits_{s'\in\mathcal B(s)} {f(s',F)}.
 \end{array}
\end{equation}
\end{proposition}
{\small
\begin{proof}
Fix a frame that satisfies property \eqref{EQ:PKM6}, an arbitrary model $M$ based on it and an arbitrary state $s\in S$ and let $\circ$ be the belief change function defined by \eqref{RI}.  Let $\phi,\psi\in\Phi_0$ be in the domain of $\circ$ (that is,  $\Vert\phi\Vert\ne\varnothing$ and $\Vert\psi\Vert\ne\varnothing$) and suppose that $\psi\in K_s\circ\phi$ and $\phi\in K_s\circ\psi$. Then, $\forall s'\in\mathcal B(s)$, $f(s',\Vert\phi\Vert)\subseteq\Vert\psi\Vert$ and $f(s',\Vert\psi\Vert)\subseteq\Vert\phi\Vert$ and thus, by Property \eqref{EQ:PKM6},
\begin{equation}
\label{EQ:due}
\bigcup\limits_{s'\in\mathcal B(s)} {f(s',\Vert\phi\Vert)} =\bigcup\limits_{s'\in\mathcal B(s)} {f(s',\Vert\psi\Vert)}
\end{equation}
It follows from \eqref{EQ:due} that,  for every $\chi\in\Phi_0$, $\chi\in K_s\circ\phi$ if and only if $\chi\in K_s\circ\psi$.\footnote{
In fact, $\chi\in K_s\circ\phi$ if and only if, $\forall s'\in\mathcal B(s), f(s',\Vert\phi\Vert)\subseteq\Vert\chi\Vert$ which implies that $\bigcup\limits_{s'\in\mathcal B(s)} {f(s',\Vert\phi\Vert)} \subseteq\Vert\chi\Vert$, so that, by \eqref{EQ:due}, $\bigcup\limits_{s'\in\mathcal B(s)} {f(s',\Vert\psi\Vert)}\subseteq\Vert\chi\Vert$, which implies that, $\forall s'\in\mathcal B(s), f(s',\Vert\psi\Vert)\subseteq\Vert\chi\Vert$, so that, by \eqref{RI}, $\chi\in K_s\circ\psi$. Similarly for the case where $\chi\in K_s\circ\psi$.}\\[3pt]
Conversely, fix a frame that violates property \eqref{EQ:PKM6}. Then there exist $s\in S$ and $E,F\in 2^S\setminus\varnothing$ such that,
\begin{equation}
\label{EQ:notPKM6}
\begin{array}{l}
 \forall s'\in\mathcal B(s),\,  f(s',E)\subseteq F \text{ and }\, f(s',F)\subseteq E \\[5pt]
 \text{ and } \bigcup\limits_{s'\in\mathcal B(s)} {f(s',E)} \,\ne\,\bigcup\limits_{s'\in\mathcal B(s)} {f(s',F)}.
 \end{array}
\end{equation}
Thus either $\bigcup\limits_{s'\in\mathcal B(s)} {f(s',E)} \,\nsubseteq\,\bigcup\limits_{s'\in\mathcal B(s)} {f(s',F)}$ or $\bigcup\limits_{s'\in\mathcal B(s)} {f(s',F)} \,\nsubseteq\,\bigcup\limits_{s'\in\mathcal B(s)} {f(s',E)}$. Suppose first that $\bigcup\limits_{s'\in\mathcal B(s)} {f(s',E)} \,\nsubseteq\,\bigcup\limits_{s'\in\mathcal B(s)} {f(s',F)}$. Then there must be a $\hat s\in\mathcal B(s)$ such that
\begin{equation}
\label{EQ:first}
f(\hat s,E)\,\nsubseteq\,\bigcup\limits_{s'\in\mathcal B(s)} {f(s',F)}.
\end{equation}
Let $p,q,r\in\texttt{At}$ be atomic sentences and construct a model based on this frame where $\Vert p\Vert=E$, $\Vert q\Vert=F$ and $\Vert r\Vert=\bigcup\limits_{s'\in\mathcal B(s)} {f(s',F)}$. By \eqref{EQ:notPKM6}, $\forall s'\in\mathcal B(s)$, $f(s',\Vert p\Vert)\subseteq \Vert q\Vert$ and $f(s',\Vert q\Vert)\subseteq \Vert p\Vert$, so that, by \eqref{RI}, $q\in K\circ p$ and $p\in K\circ q$. Next we show that $K\circ p\ne K\circ q$, thus obtaining a violation of axiom \eqref{EQ:KM6}. By \eqref{EQ:first}, $f(\hat s,\Vert p\Vert)\nsubseteq\Vert r\Vert$ and thus, by \eqref{RI}, $r\notin K\circ p$. On the other hand, $\forall s''\in\mathcal B(s)$, $f(s'',\Vert q\Vert)\subseteq \bigcup\limits_{s'\in\mathcal B(s)} {f(s',\Vert q\Vert)}=\Vert r\Vert$ and thus,  by \eqref{RI}, $r\in K\circ q$. Hence $K\circ p\ne K\circ q$. The case where $\bigcup\limits_{s'\in\mathcal B(s)} {f(s',F)} \,\nsubseteq\,\bigcup\limits_{s'\in\mathcal B(s)} {f(s',E)}$ is handled similarly, by constructing a model where  $\Vert p\Vert=E$, $\Vert q\Vert=F$ and $\Vert r\Vert=\bigcup\limits_{s'\in\mathcal B(s)} {f(s',E)}$.
\end{proof}
}
Axiom $(K\diamond 7)$ is different from the other axioms of belief update, because it involves the clause ``$K$ is complete''.  In our framework, completeness of a belief set $K_s$ is a property pertaining to a model, not a property of frames.\footnote{ 
For example, consider a frame where for some state $s$, $\mathcal B(s)=\{s_1,s_2\}$. A model based on this frame where, for every atomic sentence $p\in\texttt{At}$, $s_1\models p$ if and only if $s_2\models p$ is such that $K_s$ is complete. On the other hand, a different model where, for some $p\in\texttt{At}$, $s_1\models p$ and $s_2\models\lnot p$, is such that $K_s$ is not complete (because $p\notin K_s$ and $\lnot p\notin K_s$).
}\ 
As noted above, a consistent belief set $K$ is complete if it corresponds to a single MCS (or possible world), that is, if $\llbracket K\rrbracket$ is a singleton. Thus updating a complete and consistent $K$ means updating a single possible world. This property \emph{can} be captured in a frame by having $\mathcal B(s)$ be a singleton.  This is the motivation for the following result.\smallskip
\par
Define a state $s$ in a frame to be \emph{pointed} if $\mathcal B(s)$ is a singleton. It is clear that at a pointed state $s$ in any model the belief set $K_s$ is complete.\footnote{
Fix an arbitrary model and a state $s$ and let $K_s$ the belief set at $s$. If $\mathcal B(s)=\{s'\}$ then, $\forall \phi\in\Phi_0$, either $s'\in\Vert\phi\Vert$, in which case, by definition of $K_s$, $\phi\in K_s$, or $s'\notin\Vert\phi\Vert$ (that is, $s'\in\Vert\lnot\phi\Vert$) and thus $\lnot\phi\in K_s$.
}
\begin{proposition}
\label{PROP:KM7}
Let $\mathcal F_{\diamond7}$ be the class of frames that satisfy the following property: $\forall s,s'\in S$,
\begin{equation}
\tag{$P_{\diamond 7}$}
\label{PKM7}
\text{if } \mathcal B(s)=\{s'\}  \text{ then, } \forall E,F\in 2^S,\,\, f(s',E\cup F)\subseteq f(s',E)\cup f(s',F).
\end{equation}
Then the following axiom
\begin{equation}
\tag{$\diamond 7$}
\label{KM7}
\text{if }  K \text{ is complete then } (K\circ\phi)\cap(K\circ\psi)\subseteq K\circ(\phi\vee\psi)
\end{equation}
is valid at every pointed state of every model based on a frame in $\mathcal F_{\diamond7}$ and, conversely, if a frame is not in $\mathcal F_{\diamond7}$ then the axiom is not valid on it.
\end{proposition}
{\small
\begin{proof}
Fix a frame that satisfies property \eqref{PKM7}, an arbitrary model $M$ based on it and an arbitrary pointed state $s\in S$ and let $\circ$ be the belief change function based on $K_s$  defined by \eqref{RI}. Let $s'\in S$ be such that $\mathcal B(s)=\{s'\}$. Let $\phi,\psi\in\Phi_0$ be in the domain of $\circ$ and fix an arbitrary $\chi\in (K_s\circ\phi)\cap (K_s\circ\psi)$. Then, by  \eqref{RI},   $f(s',\Vert\phi\Vert)\subseteq\Vert\chi\Vert$ and $f(s',\Vert\psi\Vert)\subseteq\Vert\chi\Vert$ and thus, by Property \eqref{PKM7} and the fact that $\Vert\phi\Vert\cup\Vert\psi\Vert=\Vert\phi\vee\psi\Vert$, $f(s',\Vert\phi\vee\psi\Vert)\subseteq\Vert\chi\Vert$, that is, by \eqref{RI}, $\chi\in K_s\circ(\phi\vee\psi)$.\\[3pt]
Conversely, fix a frame that violates property \eqref{PKM7}. Then there exist $s,s'\in S$ and $E,F\in 2^S$ such that
\begin{equation}
\label{fornegPKM7}
\begin{array}{ll}
(a)&\mathcal B(s)=\{s'\}, \text{ and}\\[3pt]
(b)&f(s',E\cup F)\nsubseteq f(s',E)\cup f(s',F).
\end{array}
\end{equation}
Let $p,q,r\in\texttt{At}$ be atomic formulas and construct a model where $\Vert p\Vert=E$, $\Vert q\Vert=F$ and $\Vert r\Vert=f(s',E)\cup f(s',F)$. Then, by  \eqref{fornegPKM7} (since $\Vert p\Vert\cup\Vert q\Vert=\Vert p\vee q\Vert$), $f(s',\Vert p\vee q\Vert)\nsubseteq\Vert r\Vert$ and thus, by \eqref{RI}, $r\notin K\circ(p\vee q)$. On the other hand, since $f(s',\Vert p\Vert)\subseteq f(s',\Vert p\Vert)\cup f(s',\Vert q\Vert)=\Vert r\Vert$ and $f(s',\Vert q\Vert)\subseteq (f(s',\Vert p\Vert)\cup f(s',\Vert q\Vert)=\Vert r\Vert$, $r\in K\circ p$ and $r\in K\circ q$, yielding a violation of axiom \eqref{KM7}, since $K_s$ is complete because $\mathcal B(s)$ is a singleton.
\end{proof}
}
The following proposition provides a similar characterization of Axiom $(K\diamond 9)$.
\begin{proposition}
\label{PROP:KM9}
Let $\mathcal F_{\diamond9}$ be the class of frames that satisfy the following property: $\forall s,s'\in S$,
\begin{equation}
\tag{$P_{\diamond 9}$}
\label{PKM9}
\begin{array}{l}
\text{if } \mathcal B(s)=\{s'\} \text{ then, } \forall E,F\in 2^S,\\[3pt]
\text{ if } f(s',E)\cap F\ne\varnothing\,\text{ then, } f(s',E\cap F)\subseteq f(s',E)\cap F.
\end{array}
\end{equation}
Then the following axiom
\begin{equation}
\tag{$\diamond 9$}
\label{KM9}
\text{if }  K \text{ is complete and } \lnot\psi\notin(K\circ\phi) \text{ then } (K\circ\phi)+\psi\subseteq K\circ(\phi\wedge\psi)
\end{equation}
is valid at every pointed state of every model based on a frame in $\mathcal F_{\diamond9}$ and, conversely, if a frame is not in $\mathcal F_{\diamond9}$ then the axiom is not valid on it.
\end{proposition}
{\small
\begin{proof}
Fix a frame that satisfies Property \eqref{PKM9}, an arbitrary model based on it, a pointed state $s$, and let $s'\in S$ be such that $\mathcal B(s)=\{s'\}$. Let  $K_s$ the belief set at $s$ and $\circ$ be the belief change function (based on $K_s$) defined by \eqref{RI}. Let $\phi,\psi\in\Phi_0$ be two formulas such that $\phi\wedge\psi$ is in the domain of $\circ$ (that is $\Vert\phi\wedge\psi\Vert\ne\varnothing$, which implies that also $\Vert\phi\Vert\ne\varnothing$ and $\Vert\psi\Vert\ne\varnothing$) and suppose that $\lnot\psi\notin K_s\circ\phi$, that is, $f(s',\Vert\phi\Vert)\cap \Vert\psi\Vert\ne\varnothing$. Then, by Property \eqref{PKM9} (and noting that $\Vert\phi\wedge\psi\Vert=\Vert\phi\Vert\cap\Vert\psi\Vert$),
\begin{equation}
\label{EQ:14a}
f(s',\Vert\phi\wedge\psi\Vert) \subseteq f(s',\Vert \phi\Vert)\cap\Vert\psi\Vert.
\end{equation}
We need to show that $Cn((K_s\circ\phi)\cup\{\psi\})\subseteq K_s\circ(\phi\wedge\psi)$. Let $\chi\in Cn((K_s\circ\phi)\cup\{\psi\})$; then, since, by (B) of Lemma \ref{LEM:KfromModel}, $K_s\circ\phi$ is deductively closed, $(\psi\rightarrow\chi)\in K_s\circ\phi$, that is, $f(s',\Vert\phi\Vert)\subseteq\Vert\psi\rightarrow\chi\Vert=(S\setminus\Vert\psi\Vert)\cup\Vert\chi\Vert$, which is equivalent to
\begin{equation}
\label{EQ:15a}
f(s',\Vert\phi\Vert)\cap\Vert\psi\Vert\subseteq \Vert\chi\Vert.
\end{equation}
It follows from \eqref{EQ:14a} and \eqref{EQ:15a} that $f(s',\Vert\phi\wedge\psi\Vert) \subseteq \Vert\chi\Vert$ so that, by \eqref{RI} (since $\mathcal B(s)=\{s'\}$), $\chi\in K\circ(\phi\wedge\psi)$.\\[4pt]
Conversely, fix a frame that violates Property \eqref{PKM9}. Then there exist $s,s'\in S$ and $E,F\in 2^S$ such that
 \begin{equation}
 \label{EQ:16a}
 \begin{array}{ll}
 \text{(a)}&\mathcal B(s)=\{s'\}\\[3pt]
 \text{(b)}&f(s',E)\cap F\ne\varnothing,\\[3pt]
 \text{(c)}&f(s',E\cap F)\nsubseteq f(s',E)\cap F.
 \end{array}
 \end{equation}
 Let $p,q,r\in\texttt{At}$ be atomic formulas and construct a model where $\Vert p\Vert=E$, $\Vert q\Vert=F$ and $\Vert r\Vert = f(s',E)\cap F$. Then, by (b) of \eqref{EQ:16a}, $f(s',\Vert p\Vert)\cap\Vert q\Vert\ne\varnothing$ so that $\lnot q\notin K_s\circ p$. Furthermore, by (c)  of \eqref{EQ:16a}, $f(s',\Vert p\wedge q\Vert)\nsubseteq\Vert r\Vert$ and thus $r\notin K\circ(p\wedge q)$. To obtain a violation of Axiom \eqref{KM9} it only remains to show that $r\in Cn((K\circ p)\cup\{q\})$, which is equivalent to $(q\rightarrow r)\in K\circ p$ (since, by (B) of Lemma \ref{LEM:KfromModel}, $K\circ p$ is deductively closed); that is, we have to show that $f(s',\Vert p\Vert)\subseteq\Vert q\rightarrow r\Vert=(S\setminus\Vert q\Vert)\cup\Vert r\Vert$. But this is equivalent to $f(s',\Vert p\Vert)\cap \Vert q\Vert\ne\varnothing$, which is our hypothesis (namely, (b) of \eqref{EQ:16a}).
\end{proof}
}
\smallskip
We now return to the observation made in Remark \ref{REM:aboutKM8} about axiom $(K\diamond8)$, which \cite{Pepetal96} propose as a translation of axiom $(U8)$ in \cite{KatMen91} within the context of belief sets. Let $W$ denote the set of maximally consistent sets (MCS) of formulas in $\Phi_0$ (also called possible worlds). Recall that, given a set of formulas $\Gamma\subseteq\Phi_0$, $\llbracket\Gamma\rrbracket$ denotes the set of MCS that contain all the formulas in $\Gamma$: $\llbracket\Gamma\rrbracket=\{w\in W: \forall\phi\in\Gamma, \phi\in w\}$. Recall also that $(K\diamond8)$ is the following requirement: if  $\llbracket K\rrbracket\ne\varnothing\text{ then }K\diamond\phi=\bigcap_{w\in\llbracket K\rrbracket}{(w\diamond\phi)}.$ Since we restricted attention to the case where the initial belief set $K$ is consistent, the clause $\llbracket K\rrbracket\ne\varnothing$ is superfluous. Thus "axiom" $(K\diamond8)$ can be stated as:
$$K\diamond\phi=\bigcap_{w\in\llbracket K\rrbracket}{(w\diamond\phi)}.$$
Consider the subclass of frames where $S=W$. Given any such frame, a natural model based on it is one where, $\forall w\in W, \forall\phi\in\Phi_0$, $w\models\phi$ if and only if $\phi\in w$, so that $\Vert \phi\Vert=\llbracket\phi\rrbracket$. Thus the definition of the initial belief set $K$ at $w\in W$, namely $K_w=\{\phi\in\Phi_0: \mathcal B(w)\subseteq\Vert \phi\Vert=\llbracket\phi\rrbracket\}$, implies that $\mathcal B(w)\subseteq\llbracket K_w\rrbracket$. Furthermore, if either $\mathcal B(w)$ is finite or the set of atomic formulas \texttt{At} is finite, then $\llbracket K_w\rrbracket\subseteq\mathcal B(w)$. Let us focus on this case, so that $\mathcal B(w)=\llbracket K_w\rrbracket$. For $w'\in W$ and $\phi\in\Phi_0$ define $w'\diamond\phi\subseteq\Phi_0$ as follows: $\forall \psi\in\Phi_0,$
\begin{equation}
\label{EQ:wdiamfi}
\psi\in (w'\diamond\phi)\quad\text{if and only if}\quad f(w',\llbracket\phi\rrbracket)\subseteq\llbracket\psi\rrbracket.
\end{equation}
Then (since $\mathcal B(w)=\llbracket K_w\rrbracket$)
\begin{equation}
\label{KM8inW}
  K_w\diamond\phi=\bigcap_{w'\in\llbracket K_w\rrbracket}{(w'\diamond\phi)}
\end{equation}
which is exactly $(K\diamond 8)$.\footnote{
Proof. First we show that $K_w\diamond\phi\subseteq\bigcap_{w'\in\mathcal B(w)}{ (w'\diamond\phi)}$. Fix an arbitrary $\psi\in K_w\diamond\phi$. Then, by \eqref{RI}, $\forall w'\in\mathcal B(w), f(w',\llbracket\phi\rrbracket)\subseteq\llbracket\psi\rrbracket$, that is, by \eqref{EQ:wdiamfi}, $\forall w'\in\mathcal B(w),\psi\in(w'\diamond\phi)$ so that $\psi\in\bigcap_{w'\in\mathcal B(w)}{ (w'\diamond\phi)}$. Next we show that $\bigcap_{w'\in\mathcal B(w)}{(w'\diamond\phi)}\subseteq K_w\diamond\phi$. Let $\psi\in \bigcap_{w'\in\mathcal B(w)}{(w'\diamond\phi)}$. Then, $\forall w'\in\mathcal B(w),\psi\in (w'\diamond\phi)$, that is, by \eqref{EQ:wdiamfi}, $\forall w'\in\mathcal B(w), f(w',\llbracket\phi\rrbracket)\subseteq\llbracket\psi\rrbracket$ so that, by \eqref{RI}, $\psi\in K_w\diamond\phi$.
}
\begin{definition}
\label{DEF:update_frame}
An \emph{update frame} is a frame that (besides the properties of Definition \ref{DEF:frame}) satisfies the properties of Propositions \ref{PROP:K_circ_2}, \ref{PROP:KM5/AGM7}, \ref{PROP:KM6} and \ref{PROP:KM7}, namely:
\begin{enumerate}
\item[]$(P_{\diamond2})$\quad$ \text{ if } \mathcal B(s)\subseteq E\text{ then, }\forall s'\in\mathcal B(s), f(s',E)\subseteq\mathcal B(s).$
\item[]$(P_{\diamond5/\ast7})$\quad if, $\forall s'\in \mathcal B(s), \,f(s',E\cap F)\subseteq G,$\\[3pt]
\phantom{($P_{\diamond5/\ast7}$)}\quad then, $\forall s'\in \mathcal B(s),  \,f(s',E)\cap F\subseteq G.$
\item[]$(P_{\diamond6})$\quad if,  $\forall s'\in\mathcal B(s),\,  f(s',E)\subseteq F \text{ and }\, f(s',F)\subseteq E$ \\[3pt]
 \phantom{($P_{\diamond6}$)}\quad then  $\bigcup\limits_{s'\in\mathcal B(s)} {f(s',E)} =\bigcup\limits_{s'\in\mathcal B(s)} {f(s',F)}.$
 \item[]$(P_{\diamond7})$\quad if $\mathcal B(s)=\{s'\}$ then,  $\forall E,F\in 2^S$,\\[3pt]
 \phantom{$(P_{\diamond7})$}\quad $f(s',E\cup F)\subseteq f(s',E)\cup f(s',F).$
\end{enumerate}
We denote by $\mathcal F_{\mathlarger{\diamond}}$ the class of such frames.
\end{definition}
For  every model based on a frame in $\mathcal F_{\mathlarger{\diamond}}$ and for every state $s$ in that model, the belief change function (based on $K_s=\{\phi\in\Phi_0:\mathcal B(s)\subseteq \Vert\phi\Vert\}$) defined by \eqref{RI} satisfies the following axioms for belief update:
\begin{enumerate}
 \setcounter{enumi}{0}
\item $(K\diamond0)$ (by (B) of Lemma \ref{LEM:KfromModel}),
\item $(K\diamond1)$ (by (2) of Proposition \ref{PROP:validaxioms}),
\item $(K\diamond2)$ (by Proposition \ref{PROP:K_circ_2}),
\item $(K\diamond4)$ (by (3) of Proposition \ref{PROP:validaxioms}),
\item $(K\diamond5)$ (by Proposition \ref{PROP:KM5/AGM7}),
\item $(K\diamond6)$ (by Proposition \ref{PROP:KM6}),
\item $(K\diamond7)$ (by Proposition \ref{PROP:KM7}).
\end{enumerate}
The only remaining axiom is $K\diamond3$ which says that $K\diamond \phi=\Phi_0$ if and only if $\phi$ is a contradiction. When $\phi$ is a contradiction, then $\Vert\phi\Vert=\varnothing$ in every model and thus $K\circ\phi$ is not defined according to \eqref{RI}. This is easily fixed by adding the stipulation that $K\circ\phi=\Phi_0$ when $\phi$ is a contradiction. However, in a given model there may also be consistent formulas $\phi$ such that $\Vert\phi\Vert=\varnothing$, so that - again - $K\circ\phi$ is not defined according to \eqref{RI}. Thus the question arises as to whether any partial belief change function $\circ$ defined by \eqref{RI} can be extended to a full-domain belief update function $\diamond$ (Definition \ref{DEF:update}). Conversely, it is natural to ask whether any  full-domain belief update function coincides with the belief change function defined by \eqref{RI} in some model. The following proposition answers both questions in the affirmative. Since the proof of Proposition \ref{PROP:char_update} is rather lengthy it is relegated to the appendix.
\begin{proposition}
\label{PROP:char_update}
The class $\mathcal F_{\mathlarger{\diamond}}$ of update frames characterizes the set of full-domain belief update functions in the following sense:
\begin{enumerate}
\item[(A)] For  every model based on a frame in $\mathcal F_{\mathlarger{\diamond}}$ and for every state $s$ in that model, the belief change function $\circ$ (based on $K_s=\{\phi\in\Phi_0:\mathcal B(s)\subseteq \Vert\phi\Vert\}$) defined by \eqref{RI} can be extended to a full-domain belief update function $\diamond$ (Definition \ref{DEF:update}).
\item[(B)] Let $K\subset \Phi_0$ be a consistent and deductively closed set and let $\diamond:\Phi_0\to 2^{\Phi_0}$ be a belief update function based on $K$ (Definition \ref{DEF:update}). Then there exists a frame in $\mathcal F_{\mathlarger{\diamond}}$, a model based on that frame and a state $s$ in that model such that (1) $K=K_s=\{\phi\in\Phi_0:\mathcal B(s)\subseteq \Vert\phi\Vert\}$ and (2) the partial belief change function $\circ$ (based on $K_s$) defined by \eqref{RI}  is such that $K_s\circ\phi=K_s\diamond\phi$ for every consistent formula $\phi$.
\end{enumerate}
\end{proposition}
\begin{remark}
\label{REM:char4diamond9}
As noted in Remark \ref{REM:KM9}, in \cite{KatMen91} Katsuno and Mendelzon also put forward a stronger notion of belief update, obtained by replacing (in our translation) axioms $(K\diamond6)$ and $(K\diamond7)$ with axiom $(K\diamond9)$. Proposition  \ref{PROP:char_update9} below shows that a result analogous to the characterization of Proposition \ref{PROP:char_update} applies also to this stronger notion. Since the proof is very similar to the proof of Proposition \ref{PROP:char_update} we omit it.
\end{remark}
\begin{definition}
\label{DEF:strong-update-frame}
A \emph{strong update frame} is a frame that (besides the properties of Definition \ref{DEF:frame}) satisfies the following properties:
\begin{enumerate}
\item[]$(P_{\diamond2})$\quad if $\mathcal B(s)\subseteq E\text{ then, }\forall s'\in\mathcal B(s), f(s',E)\subseteq\mathcal B(s).$
\item[]$(P_{\diamond5/\ast7})$\quad if, $\forall s'\in \mathcal B(s), \,f(s',E\cap F)\subseteq G,$\\[3pt]
\phantom{($P_{\diamond5/\ast7}$)}\quad then, $\forall s'\in \mathcal B(s),  \,f(s',E)\cap F\subseteq G.$
\item[]$(P_{\diamond9})$\quad if $\mathcal B(s)=\{s'\}$  then,  $\forall E,F\in 2^S,$ if $f(s',E)\cap F\ne\varnothing$\\[3pt]
\phantom{($P_{\diamond9})$}\quad then, $ f(s',E\cap F)\subseteq f(s',E)\cap F$.
\end{enumerate}
We denote by $\mathcal F_{\mathlarger{\diamond\diamond}}$ the class of such frames.
\end{definition}
\begin{proposition}
\label{PROP:char_update9}
The class $\mathcal F_{\mathlarger{\diamond\diamond}}$ of strong update frames characterizes the set of strong belief update functions (Definition \ref{DEF:strong_update function}):
\begin{enumerate}
\item[(A)] For  every model based on a frame in $\mathcal F_{\mathlarger{\diamond\diamond}}$ and for every state $s$ in that model, the belief change function $\circ$ (based on $K_s=\{\phi\in\Phi_0:\mathcal B(s)\subseteq \Vert\phi\Vert\}$) defined by \eqref{RI} can be extended to a full-domain belief update function $\diamond$ that satisfies axioms $(K\diamond0)$-$(K\diamond5)$ and $(K\diamond9)$.
\item[(B)] Let $K\subset \Phi_0$ be a consistent and deductively closed set and let $\diamond:\Phi_0\to 2^{\Phi_0}$ be a belief update function based on $K$ that satisfies axioms $(K\diamond0)$-$(K\diamond5)$ and $(K\diamond9)$. Then there exists a frame in $\mathcal F_{\mathlarger{\diamond\diamond}}$, a model based on that frame and a state $s$ in that model such that (1) $K=K_s=\{\phi\in\Phi_0:\mathcal B(s)\subseteq \Vert\phi\Vert\}$ and (2) the partial belief change function $\circ$ (based on $K_s$) defined by \eqref{RI}  is such that $K_s\circ\phi=K_s\diamond\phi$ for every consistent formula $\phi$.
\end{enumerate}
\end{proposition}
 \section{Frame correspondence. Part 2: Revision}
 \label{SEC:CorrespondenceRevision}
 As noted in Remark \ref{REMcompare-circ-diamond}, axioms $(K\ast4)$ and $(K\ast8)$ can be viewed as the crucial axioms that distinguish AGM belief revision from the strong version of KM belief update. In this section we provide a semantic characterization of these two axioms and a characterization of AGM belief revision functions along the lines of that provided for belief update (Propositions \ref{PROP:char_update} and  \ref{PROP:char_update9}), while in Section \ref{SEC:comparison} we compare belief revision and belief update by focusing on the interpretation of the semantic properties corresponding to $(K\ast4)$ and $(K\ast8)$ in relation to the semantic properties corresponding to $(K\diamond2)$ and $(K\diamond9)$.
 \begin{proposition}
\label{PROP:AGM4}
The following axiom:
\begin{equation}
\tag{$\ast4$}
\label{AGM4}
\text{if } \lnot\phi\notin K\text{ then }K\subseteq K\circ\phi
\end{equation}
is characterized by the following property of frames: $\forall s\in S,\forall E\subseteq S$,
\begin{equation}
\tag{$P_{\ast4}$}
\label{PAGM4}
 \text{ if } \mathcal B(s)\cap E\ne\varnothing\text{ then, }\forall s'\in\mathcal B(s), f(s',E)\subseteq\mathcal B(s)\cap E.
\end{equation}
\end{proposition}
{\small
\begin{proof}
Fix a frame that satisfies Property \eqref{PAGM4}, an arbitrary model based on it, a state $s$ and let $K_s$ the belief set at $s$ and $\circ$ be the belief change function (based on $K_s$) defined by \eqref{RI}. Let $\phi\in\Phi_0$ be a formula in the domain of $\circ$ and suppose that $\lnot\phi\notin K_s$, that is, $\mathcal B(s)\nsubseteq\Vert\lnot\phi\Vert$, i.e. $\mathcal B(s)\cap\Vert\phi\Vert\ne\varnothing$. We need to show that $K_s\subseteq K_s\circ\phi$. Fix an arbitrary $\psi\in K_s$; then $\mathcal B(s)\subseteq\Vert\psi\Vert$. Since $\mathcal B(s)\cap\Vert\phi\Vert\ne\varnothing$, by Property \eqref{PAGM4} (with $E=\Vert\phi\Vert$) $\forall s'\in\mathcal B(s), f(s',\Vert\phi\Vert)\subseteq\mathcal B(s)\cap \Vert\phi\Vert$. Thus (since $\mathcal B(s)\cap\Vert\phi\Vert\subseteq\mathcal B(s)\subseteq\Vert\psi\Vert$),
 $\forall s'\in\mathcal B(s), f(s',\Vert\phi\Vert)\subseteq\Vert\psi\Vert$, that is, $\psi\in K\circ\phi$.\\[3pt]
Conversely, consider a frame that violates Property \eqref{PAGM4}. Then there exist $s\in S$ and $E\subseteq S$ such that $\mathcal B(s)\cap E\ne\varnothing$, $s'\in\mathcal B(s)$ and $f(s',E)\nsubseteq \mathcal B(s)\cap E$. Let $p,q\in\texttt{At}$ and construct a model based on this frame where $\Vert p\Vert=E$ and $\Vert q\Vert=\mathcal B(s)\cap E$. Since $s'\in\mathcal B(s)$ and  $\mathcal B(s)\cap \Vert p\Vert\ne\varnothing$, $\lnot p\notin K_s$. Since $f(s',\Vert p\Vert)\nsubseteq\mathcal B(s)\cap E=\Vert q\Vert$,
\begin{equation}
\label{EQ:10}
q\notin K_s\circ p.
\end{equation}
By (2) of Proposition \ref{PROP:validaxioms}, $p\in K\circ p$. Thus, for every formula $\phi$, $(p\rightarrow\phi)\in K\circ p$ if and only if $\phi\in K\circ p$ (since, by (B) of Lemma \ref{LEM:KfromModel}, $K\circ p$ is deductively closed). Hence, by \eqref{EQ:10},
\begin{equation}
\label{EQ:11}
(p\rightarrow q)\notin K_s\circ p.
\end{equation}
Next we show that $(p\rightarrow q)\in K_s$ so that $K_s\nsubseteq K_s\circ p$, yielding a violation of axiom \eqref{AGM4} (since $\lnot p\notin K_s$). We need to show that $\mathcal B(s)\subseteq\Vert p\rightarrow q\Vert=\Vert\lnot p\Vert\cup\Vert q\Vert=(S\setminus\Vert p\Vert)\cup\Vert q\Vert$. First note that
\begin{equation}
\label{EQ:12}
\mathcal B(s)=(\mathcal B(s)\cap(S\setminus E))\,\cup\, (\mathcal B(s)\cap E).
\end{equation}
Since $E=\Vert p\Vert$, $B(s)\cap(S\setminus E)=B(s)\cap\Vert\lnot p\Vert\subseteq\Vert\lnot p\Vert\subseteq\Vert\lnot p\Vert)\cup\Vert q\Vert=\Vert p\rightarrow q\Vert$. Thus
\begin{equation}
\label{EQ:13}
B(s)\cap(S\setminus E)\subseteq\Vert p\rightarrow q\Vert.
\end{equation}
Since $\mathcal B(s)\cap E=\Vert q\Vert$, $\mathcal B(s)\cap E\subseteq\Vert\lnot p\Vert\cup\Vert q\Vert=\Vert p\rightarrow q\Vert$. It follows from this, \eqref{EQ:13} and \eqref{EQ:12} that $\mathcal B(s)\subseteq\Vert p\rightarrow q\Vert$.
\end{proof}
}
\begin{proposition}
\label{PROP:AGM8}
The following axiom:
\begin{equation}
\tag{$\ast8$}
\label{AGM8}
\text{if }\, \lnot\psi\notin K\circ\phi\text{ then }\, K\circ\phi+\psi\subseteq K\circ(\phi\wedge\psi)
\end{equation}
is characterized by the following property of frames: $\forall s\in S,\forall E,F\in 2^S$,
\begin{equation}
\tag{$P_{\ast8}$}
\label{PAGM8}
\begin{array}{l}
 \text{ if }\, \exists \hat s\in\mathcal B(s) \text{ such that }f(\hat s,E)\cap F\ne\varnothing  \text{ then,  } \\[7pt]
\forall s'\in\mathcal B(s),\, f(s',E\cap F)\,\,{\mathlarger{\mathlarger \subseteq}} \bigcup\limits_{s''\in\mathcal B(s)} \left(f(s'',E)\cap F\right).
 \end{array}
 \end{equation}
\end{proposition}
{\small
\begin{proof}
Fix a frame that satisfies Property \eqref{PAGM8}, an arbitrary model based on it, a state $s$, and let $K_s$ the belief set at $s$ and $\circ$ be the belief change function (based on $K_s$) defined by \eqref{RI}. Let $\phi,\psi\in\Phi_0$ be two formulas such that $\phi\wedge\psi$ is in the domain of $\circ$ (that is $\Vert\phi\wedge\psi\Vert\ne\varnothing$, which implies that also $\Vert\phi\Vert\ne\varnothing$ and $\Vert\psi\Vert\ne\varnothing$) and suppose that $\lnot\psi\notin K_s\circ\phi$, that is, there exists an $\hat s\in\mathcal B(s)$  such that $f(\hat s,\Vert\phi\Vert)\cap \Vert\psi\Vert\ne\varnothing$. Then, by Property \eqref{PAGM8} (and noting that $\Vert\phi\wedge\psi\Vert=\Vert\phi\Vert\cap\Vert\psi\Vert$),
\begin{equation}
\label{EQ:14}
\forall s'\in\mathcal B(s),\,f(s',\Vert\phi\wedge\psi\Vert)\,\, \subseteq\,\, \bigcup\limits_{s''\in \mathcal B(s)} \left(f(s'',\Vert\phi\Vert)\cap \Vert\psi\Vert\right).
\end{equation}
We need to show that $Cn((K_s\circ\phi)\cup\{\psi\})\subseteq K_s\circ(\phi\wedge\psi)$. Fix an arbitrary $\chi\in Cn((K_s\circ\phi)\cup\{\psi\})$; then, since, by (B) of Lemma \ref{LEM:KfromModel}, $K_s\circ\phi$ is deductively closed, $(\psi\rightarrow\chi)\in K_s\circ\phi$, that is,
\begin{equation*}
\forall s'\in\mathcal B(s),\, f(s',\Vert\phi\Vert)\subseteq\Vert\psi\rightarrow\chi\Vert=(S\setminus\Vert\psi\Vert)\cup\Vert\chi\Vert,
\end{equation*}
which is equivalent to
\begin{equation*}
\forall s'\in\mathcal B(s),\, f(s',\Vert\phi\Vert)\cap\Vert\psi\Vert\subseteq\Vert\chi\Vert,
\end{equation*}
so that
\begin{equation}
\label{EQ:15}
\bigcup\limits_{s'\in \mathcal B(s)} \left(f(s',\Vert\phi\Vert)\cap \Vert\psi\Vert\right)\subseteq\Vert\chi\Vert.
\end{equation}
It follows from \eqref{EQ:14} and \eqref{EQ:15} that $\forall s'\in\mathcal B(s),f(s',\Vert\phi\wedge\psi\Vert)\subseteq\Vert\chi\Vert$, that is, $\chi\in K\circ(\phi\wedge\psi)$.\\[4pt]
 Conversely, fix a frame that violates Property \eqref{PAGM8}. Then there exist $s,s_1,s_2\in S$ and $E,F\in 2^S$ such that
 \begin{equation}
 \label{EQ:18}
 \begin{array}{ll}
 \text{(a)}&s_1\in\mathcal B(s)\text{ and }f(s_1,E)\cap F\ne\varnothing,\\[6pt]
 \text{(b)}&s_2\in\mathcal B(s)\text{ and }f(s_2,E\cap F)\nsubseteq\bigcup\limits_{s'\in \mathcal B(s)} \left(f(s',E)\cap F\right).
 \end{array}
 \end{equation}
 Let $p,q,r\in\texttt{At}$ be atomic formulas and construct a model where $\Vert p\Vert=E$, $\Vert q\Vert=F$ and $\Vert r\Vert = \bigcup\limits_{s'\in \mathcal B(s)} \left(f(s',E)\cap F\right)$. Then,
 \begin{equation}
 \label{EQ:19}
 \begin{array}{ll}
 \text{(A)}&\text{by (a) of \eqref{EQ:18}, } \lnot q\notin K_s\circ p\\[6pt]
 \text{(B)}&\text{by (b) of \eqref{EQ:18}, } r\notin K\circ(p\wedge q).
 \end{array}
 \end{equation}
To obtain a violation of Axiom \eqref{AGM8} it only remains to show that $r\in Cn((K\circ p)\cup\{q\})$, which is equivalent to (since, by (B) of Lemma \ref{LEM:KfromModel}, $K\circ p$ is deductively closed) $(q\rightarrow r)\in K\circ p$; that is, we have to show that $\forall s'\in\mathcal B(s)$, $f(s',\Vert p\Vert)\subseteq\Vert q\rightarrow r\Vert=(S\setminus\Vert q\Vert)\cup\Vert r\Vert$. Fix an arbitrary $s'\in\mathcal B(s)$. If $f(s',\Vert p\Vert)\cap\Vert q\Vert=\varnothing$ then $f(s',\Vert p\Vert)\subseteq(S\setminus\Vert q\Vert)\subseteq(S\setminus\Vert q\Vert)\cup\Vert r\Vert$. If $f(s',\Vert p\Vert)\cap\Vert q\Vert\ne\varnothing$ then $f(s',\Vert p\Vert)\cap\Vert q\Vert\subseteq\bigcup\limits_{s''\in \mathcal B(s)} \left(f(s'',\Vert p\Vert)\cap \Vert q\Vert\right)=\Vert r\Vert$; thus $f(s',\Vert p\Vert)\cap\Vert q\Vert\subseteq\Vert r\Vert$ which is equivalent to $f(s',\Vert p\Vert)\subseteq(S\setminus\Vert q\Vert)\cup\Vert r\Vert$.
 \end{proof}
 }
\begin{definition}
\label{DEF:revision-frame}
A \emph{revision frame} is a frame that (besides the properties of Definition \ref{DEF:frame}) satisfies the properties of Propositions \ref{PROP:AGM4} and \ref{PROP:AGM8}, namely:
\begin{enumerate}
\item[]$(P_{\ast4})$\quad if $\mathcal B(s)\cap E\ne\varnothing$ then, $\forall s'\in\mathcal B(s), f(s',E)\subseteq\mathcal B(s)\cap E$.
\item[]$(P_{\ast8})$\quad if, $\exists \hat s\in\mathcal B(s)$ such that $f(\hat s,E)\cap F\ne\varnothing$ then, \\[3pt]
\phantom{$(P_{\ast8})$}\quad$\forall s'\in\mathcal B(s),\, f(s',E\cap F)\,\,{\mathlarger{\mathlarger \subseteq}} \bigcup\limits_{s''\in\mathcal B(s)} \left(f(s'',E)\cap F\right)$.
\end{enumerate}
We denote by $\mathcal F_{\mathlarger{\ast}}$ the class of such frames.
\end{definition}
The proof of the following proposition is given in the appendix.
\begin{proposition}
\label{PROP:char_revision}
The class $\mathcal F_{\mathlarger{\ast}}$ of revision frames characterizes the set of AGM  belief revision functions (Definition \ref{DEF:revision}), in the following sense:
\begin{enumerate}
\item[(A)] For  every model based on a frame in $\mathcal F_{\mathlarger{\ast}}$ and for every state $s$ in that model, the belief change function $\circ$ (based on $K_s=\{\phi\in\Phi_0:\mathcal B(s)\subseteq \Vert\phi\Vert\}$) defined by \eqref{RI} can be extended to a full-domain belief revision function $\ast$ that satisfies the AGM axioms $(K\ast1)$-$(K\ast8)$.
\item[(B)] Let $K\subset \Phi_0$ be a consistent and deductively closed set and let $\ast:\Phi_0\to 2^{\Phi_0}$ be a belief revision function based on $K$ that satisfies the AGM axioms $(K\ast1)$-$(K\ast8)$. Then there exists a frame in $\mathcal F_{\mathlarger{\ast}}$, a model based on that frame and a state $s$ in that model such that (1) $K=K_s=\{\phi\in\Phi_0:\mathcal B(s)\subseteq \Vert\phi\Vert\}$ and (2) the partial belief change function $\circ$ (based on $K_s$) defined by \eqref{RI}  is such that $K_s\circ\phi=K_s\ast\phi$ for every consistent formula $\phi$.
\end{enumerate}
\end{proposition}
 \section{Comparing KM update and AGM revision}
 \label{SEC:comparison}
 As noted in Remark \ref{REMcompare-circ-diamond}, AGM axiom $(K\ast4)$ can be seen as a strengthening of KM axiom $(K\diamond2)$. This is clear if we compare the corresponding semantic properties (Propositions \ref{PROP:K_circ_2} and \ref{PROP:AGM4}):
 \begin{enumerate}
 \item[\emph{(i)}] For $(K\diamond2)$:  if  $\mathcal B(s)\subseteq E$  then, $\forall s'\in\mathcal B(s), f(s',E)\subseteq\mathcal B(s)$.
 \item[\emph{(ii)}] For $(K\ast4)$: if $\mathcal B(s)\cap E\ne\varnothing$  then, $\forall s'\in\mathcal B(s), f(s',E)\subseteq\mathcal B(s)\cap E$.
 \end{enumerate}
 By definition of frame (Definition \ref{DEF:frame}), since $\mathcal B$ is serial, $\mathcal B(s)\ne\varnothing$ and thus $\mathcal B(s)\subseteq E$ implies that $\mathcal B(s)\cap E=\mathcal B(s)\ne\varnothing$. Hence if \emph{(i)} is satisfied then so is \emph{(ii)}. What \emph{(i)} says is that if, initially, the agent believes event $E$ then, conditional on $E$, he continues to believe everything that he believed initially. On the other hand, \emph{(ii)} says that if, among the states that the agent initially considered possible, there are states where event $E$ is true, then those states should be given priority when conditioning on $E$. In other words, when looking for $E$-states that are closest to a state that is initially considered possible, the agent should -- if possible -- first consider those states in his initial belief set  that are already $E$-states. We shall call this requirement \emph{Doxastic Priority 1} (DP1). We can restate Doxastic Priority 1 as follows:
 \begin{equation}
 \tag{DP1}
 \label{DP1}
 \begin{array}{l}
 E\text{-states in }\mathcal B(s) \text{ are to be selected as nearer to states in }\mathcal B(s)  \\
 \text{than any }E\text{-states outside of } \mathcal B(s).
 \end{array}
 \end{equation}
 This principle is reminiscent of a principle put forward by Stalnaker \cite{Stal75} in his theory of context-dependent indicative conditionals: if an indicative conditional is being evaluated at a world in the context set, then the world selected must, if possible, be within the context set as well. In our semantic framework, the set $\mathcal B(s)$ can be viewed as playing a role similar to the role played by the context set in Stalnaker's theory of indicative conditionals.
 \par
 In Remark \ref{REMcompare-circ-diamond} it was also noted that axiom $(K\ast8)$ can be seen as a strengthening of KM axiom $(K\diamond9)$. Once again, this is clear if we compare the corresponding semantic properties (Propositions \ref{PROP:KM9} and \ref{PROP:AGM8}):
 \begin{enumerate}
 \item[(I)] For $(K\diamond9)$:  if  $\mathcal B(s)=\{s'\}$  then, $\forall E,F\in 2^S$, if  $f(s',E)\cap F\ne\varnothing$ then $f(s',E\cap F)\subseteq f(s',E)\cap F$.
 \item[(II)] For $(K\ast8)$: if $\exists \hat s\in\mathcal B(s)$  such that $f(\hat s,E)\cap F\ne\varnothing$  then,  $\forall s'\in\mathcal B(s),\, f(s',E\cap F)\,\,{\mathlarger \subseteq} \bigcup\limits_{s''\in\mathcal B(s)} \left(f(s'',E)\cap F\right)$.
 \end{enumerate}
 It is clear that if $\mathcal B(s)$ is a singleton then (II) reduces to (I). Property (II) plays a similar role to (DP1): it says that if there are $E$-states closest to states in $\mathcal B(s)$ which are also $F$-states then the closest $E$-and-$F$-states to states in $\mathcal B(s)$ must be among those. We call this principle \emph{Doxastic Priority 2} (DP2):
  \begin{equation}
 \tag{DP2}
 \label{DP2}
 \begin{array}{l}
 \text{When looking for } E\text{-and-}F\text{ states outside of }\mathcal B(s), \text{ the} \\
 \text{closest }E\text{-states} \text{ that are also } F\text{-states are to be selected} \\
\text{as nearer to  }\mathcal B(s) \text{ than any }E\text{-states outside of } \mathcal B(s)\text{ that}   \\
\text{are not also }F\text{-states}.
 \end{array}
 \end{equation}
 The examples given by Katsuno and Mendelzon in \cite{KatMen91} show that (DP1) and (DP2) are not requirements that one would want to impose, in general, on a theory of belief updating. However, the fact remains that if one strengthens the definition of strong update frame (Definition \ref{DEF:strong-update-frame}) by replacing (1) $(P_{\diamond2})$ with the stronger $(P_{\ast4})$ and (2) $(P_{\diamond9})$ with the stronger $(P_{\ast8})$, then one obtains a revision frame (Definition \ref{DEF:revision-frame}); in other words, \emph{the set of revision frames is a subset of the set of strong update frames}.  This fact suggests that, as Peppas et al. point out in \cite[p. 98]{Pepetal96} "revision functions are nothing but a special kind of update operator"; indeed, they prove  (by making use of a different semantics from the one considered in this paper) that, given a consistent belief set $K$, for every revision function $\ast$ there exists an update operator $\diamond$ such that $K\ast\phi=K\diamond\phi,\,\forall\phi\in\Phi_0$ \cite[Theorem 2, p. 98]{Pepetal96}.

 \section{Related literature}
\label{SEC:literature}
The characterization results of Propositions \ref{PROP:char_update} and \ref{PROP:char_revision} provide an interpretation of $\psi\in K\diamond\phi$ in the KM framework (that is, believing $\psi$ after \emph{updating} by $\phi$) and $\psi\in K\ast\phi$ in the AGM framework (that is, believing $\psi$ after \emph{revising} by $\phi$) in terms of believing the conditional "if $\phi$ is, or were, the case, then $\psi$ is, or would be, the case".
\par
That the notion of belief update is closely related to conditionals has been pointed out before in the literature. Grahne (\cite{Gra98} considers a modal logic containing two bi-modal operators: the conditional operator > and the update operator $\diamond$ (Grahne uses the symbol $\circ$ but for continuity with our previous notation we have changed it to $\diamond$). The proposed axioms involve only the conditional operator >, while the update operator enters via two rules of inference, which Grahne calls "Ramsey's Rules" (RR):
\begin{equation}
\tag{RR}
\label{GrahneRR}
\frac{\chi\rightarrow(\phi>\psi)}{(\chi\diamond\phi)\rightarrow\psi}\quad\text{and}\quad
\frac{(\chi\diamond\phi)\rightarrow\psi}{\chi\rightarrow(\phi>\psi)}.
\end{equation}
Grahne (\cite[p. 97]{Gra98} offers the following explanation for \eqref{GrahneRR}.
\begin{quote}
The intuitive interpretation of \eqref{GrahneRR} is as follows. Let original belief state be $\chi$ and let $\phi>\psi$ stand for `If $\phi$, then $\psi$.' If $\phi>\psi$ is accepted in state $\chi$ it means that $\chi\rightarrow(\phi>\psi)$ is a theorem. Now `the minimal change' of $\chi$ `needed to accept' $\phi$, which is represented by $\chi\diamond\phi$ `also requires accepting' $\psi$, since $(\chi\diamond\phi)\rightarrow\psi$ is a theorem, according to \eqref{GrahneRR}.
\end{quote}
He also notes that, without \eqref{GrahneRR}, his logic coincides with Lewis' logic VCU for counterfactuals (\cite{Lew73}). Grahne proves that his proposed logic is sound and complete with respect to the standard semantics based on possible worlds (\cite{Gro88,Lew73}) and  concludes that G\"aerdenfor's Triviality Theorem (see below) applies only to revision operators, not to update operators.
\par
Ryan and Schobbens (\cite{RyaSch97}) point out a link between the theory of updates, the theory of counterfactuals and classical modal logic: they show that update is a classical existential modality, counterfactual is a classical universal modality and the accessibility relations corresponding to these modalities are inverses of each other. The argue that the Ramsey Rule (see \eqref{EQ:Ramsey Test} below) is simply an axiomatisation of this inverse relationship.
\par
In the belief revision literature, the attempt to relate belief revision to conditionals led to G\"{a}rdenfors' Triviality Theorem (\cite{Gae86}). In his approach the Boolean propositional language $\Phi_0$ is extended to a modal language, call it  $\Phi_>$, which includes conditionals of the form $\phi>\psi$ (if $\phi$ were the case then $\psi$ would be the case). In this extended language, belief sets are allowed to contain conditionals and indeed it is postulated \cite[p.84]{Gae86} that
\begin{equation}
\label{EQ:Ramsey Test}
\tag{R}
(\phi>\psi)\in K \text{ if and only if } \psi\in K\ast\phi,
\end{equation}
which is intended to capture the "Ramsey test" (see Footnote \ref{FT:Ramsey}). G\"{a}rdenfors proved that if \eqref{EQ:Ramsey Test} is added to the AGM postulates for belief revision, only trivialized revision operators (or revision operators defined on a trivialized set of belief sets) are allowed.\footnote{
G\"{a}rdenfors' triviality result gave rise to a sizeable literature; see, for example, \cite{Lei10,Lin96,LinRab98,LinRab92}.
}\ Note that G\"{a}rdenfors' triviality result does not apply to our framework because --   as in the original AGM theory -- we restricted the analysis to a propositional language containing only Boolean formulas.
\par
In a series of related papers, Giordano et al. (\cite{Gioetal98,Gioetal01,Gioetal05}) establish a connection between AGM belief revision and conditionals. They consider a modal language obtained by adding to a propositional logic the conditional operator >, with the usual interpretation of $\phi>\psi$ as "if $\phi$ were the case, then $\psi$ would be the case". On the semantic side they consider a selection function $f$ that takes as input a possible world $w$ and a formula $\phi$ and returns as output a set $f(w,\phi)$ of possible worlds. They don't postulate a belief relation, but instead they extract the initial beliefs from the selection function as follows: letting $\top$ denote any tautology, they define "$\phi$ is initially believed" as $f(w,\top)\subseteq\Vert\phi\Vert$. They correspondingly define the relation $R\subseteq W\times W$ as follows: $(w,w')\in R$ if and only if $w'\in f(w,\top)$ and impose axioms in their logic that make $R$ reflexive and euclidean, that is, a partition, thus imposing the logic S5 on beliefs. They also impose the condition (which they call BEL) that a conditional $\phi>\psi$ is true at world $w$ if and only if it is true at every $w'$ such that $(w,w')\in R$. They then prove the following representation result: (1) each AGM belief revision system corresponds to a model of their logic and (2) every model of their logic that satisfies a strong condition, which they call the "covering condition", determines an AGM belief revision system. A model satisfies the covering condition if, for every consistent formula $\phi$, $\Vert\phi\Vert\ne\varnothing$. Translating their semantics into our framework would require us to define $\mathcal B(s)=f(s,S)$ and to impose the following restriction on the selection function: if $s'\in f(s,S)$ then, for every $E\subseteq S$,  $f(s',E)=f(s,E)$. This approach leaves to be desired: first of all, it rules out incorrect beliefs and, secondly, it rules out the possibility that for $s',s''\in\mathcal B(s)$ the closest $E$-states to $s'$ might be different from the closest $E$-states to $s''$. A distinguishing feature of our semantics is that it allows for "small" models where it is possible for a consistent formula $\phi$ to be such that $\Vert\phi\Vert=\varnothing$. In order to prove their representation result, Giordano et al. rule this out via their "covering condition". By contrast, we are able to allow for the possibility that, in a model, $\Vert\phi\Vert=\varnothing$ even if $\phi$ is consistent, because we frame the analysis in terms of \emph{partial} belief revision functions and formulate the characterization problem in terms of the existence of an AGM/KM extension of a given partial belief change function. It should also be noted that the analysis of Giordano et al. is restricted to belief revision and does not deal with belief update.
\par
The semantics given in Section \ref{SEC:Frames} -- namely a Lewis selection function with the addition of a Kripke belief relation -- was also implicitly considered in \cite{Lei07} who proposed a different interpretation of $\psi\in K\ast\phi$:
\begin{quote}
  Let $w$ be a possible world in which the agent's actual belief set is $K$. Now consider the set $W^\prime$ of worlds $w^\prime$ in which our agent believes $\phi$ [...] we can thus reformulate the semantics of the revision operator as follows: "$\psi\in K{\Large*}\phi$" is true in $w$ if and only if all those worlds $w^{\prime\prime}$ among the members of $W^\prime$ that are maximally similar to $w$ in Lewis' sense are worlds in which the agent believes $\psi$. [...] a rational agent has a conditional belief in $\psi$ given $\phi$ if and only if: \emph{if she believed $\phi$, then she would believe $\psi$}. \cite[p. 121]{Lei07}
\end{quote}
\noindent
Thus Leitgeb suggests a very different interpretation from ours. He argues that "there are two different types of beliefs of "conditional character": \emph{beliefs in conditionals} and \emph{conditional beliefs}." \cite[p. 115]{Lei07}. We focused on the former while Leitgeb opted for the latter. Leitgeb offers several arguments in favor of his suggested interpretation, but does not establish an exact correspondence between AGM belief revision functions and the semantics he has in mind. Our objective is not to ague that our proposed interpretation is "the correct" one, but simply to show that it works, in the sense that it provides a semantic characterization, not only of KM belief update, but also of AGM belief revision that  is different from the standard one based on plausibility pre-orders. Proposition \ref{PROP:char_revision} shows that, in the  AGM theory, $\psi\in K\ast\phi$ can be consistently interpreted as belief in the conditional "if $\phi$ is (were) the case, then $\psi$ is (would be) the case".
\par
A semantics consisting of a Stalnaker selection function\footnote{
The difference between a Stalnaker selection function and a Lewis selection function is that the former requires $f(s,E)$ to be a singleton, that is, that there be a unique $E$-state closest to $s$.
}\,
 augmented with a belief relation was considered recently by G{\"u}nther and Sisti in \cite{GunSis22} who dubbed it "Stalnaker's Ramsey Test". However, the focus of \cite{GunSis22} is very different from ours: the authors do no establish a link to the AGM theory of belief revision and do not view the proposed semantics as an alternative characterization of AGM belief revision. The main purpose of \cite{GunSis22} is to argue that the "Stalnaker Ramsey Test" provides an alternative way of capturing Ramsey's inferential account, which was framed in terms of variable hypotheticals.\footnote{
A variable hypothetical is a subjective rule that Ramsey expresses as "If I meet a $\phi$ I shall regard it as a $\psi$" \cite[p. 241]{Ram50}. G{\"u}nther and Sisti \cite[p.29]{GunSis22} argue that the belief in the variable hypothetical $\forall x\left(\phi(x) \rightarrow \psi(x)\right)$ can be faithfully translated into Stalnaker semantics as follows: for all worlds the agent cannot exclude to be the actual, the most similar $\phi$-world is a $\psi$-world.
}
\par
The interpretation of both $\psi\in K\diamond\phi$ (update) and $\psi\in K\ast\phi$ (revision) suggested in this paper as belief in the conditional "if $\phi$ is (were) the case, then $\psi$ is (would be) the case" does not make a distinction between indicative and subjunctive conditionals. The view that a closest-world semantics is appropriate for both indicative and subjunctive conditionals has been defended by several authors (\cite{Dav79,Lyc01,Wea01,Nol03,Stal75}).
\par

\section{Conclusion}
\label{SEC:Conclusion}
We provided a characterization of KM belief update and AGM belief revision in terms of a semantics that consists of a selection function together with a belief relation. We have shown that the set of KM/AGM belief update/revision functions corresponds to the set of functions that can be obtained from the class of models that we considered, by identifying the initial belief set $K$ with the set of formulas that the agent believes at a state $s$ ($K=\{\phi\in\Phi_0:\mathcal B(s)\subseteq\Vert\phi\Vert\}$) and by identifying the updated belief set ($K\diamond\phi$), or the revised belief set ($K\ast\phi$), in response to input $\phi$, with the set of formulas that are the consequent of conditionals that are believed at state $s$ and have $\phi$ as antecedent. Thus our analysis shows that both belief update and belief revision can be understood in terms of belief in the conditional "if $\phi$ is (or were) the case then $\psi$ is (or would be) the case". The difference between (strong) update and revision boils down to two stronger properties (properties of doxastic priority DP1 and DP2) that are required for belief revision but not for belief update. Since every revision frame is also an update frame, our analysis confirms Peppas et al.'s assessment that, for a \emph{fixed} initial belief set $K$, "revising $K$ is much the same as updating $K$" (\cite[p. 95]{Pepetal96}.
\par
A natural next step is to go beyond the propositional language considered in this paper and study a modal language that includes a unimodal belief operator $\mathbb B$, corresponding to the belief relation $\mathcal B$, and a bimodal conditional operator $>$ corresponding to the selection function $f$.  The investigation of the corresponding modal logic and its potential use in modeling belief change is beyond the scope of this paper and is pursued in a companion paper (work in progress).
 \small
 \appendix
 \section{Proof of Propositions \ref{PROP:char_update} and \ref{PROP:char_revision}}
\textbf{Proof of Proposition \ref{PROP:char_update}.} In what follows we write $\vdash\phi$ to denote that $\phi$ is a tautology.\\
 \textbf{(A)} We need to show that the partial belief change function obtained at a state of an arbitrary model based on an update frame, can be extended to a full-domain belief update function. \emph{The purpose here is not to define the most natural extension, but to show that at least one such an extension exists.} Thus we will do so in the simplest possible way.\\
 Fix an arbitrary frame $F\in\mathcal F_{\mathlarger{\diamond}}$, an arbitrary model based on $F$,  an arbitrary state $s$ and let $K_s=\{\phi\in\Phi_0:\mathcal B(s)\subseteq\Vert\phi\Vert\}$. Let $\circ$ be the belief change function based on $K_s$ defined by \eqref{RI}, that is, $\psi\in K_s\circ\phi$ if and only if, $\Vert\phi\Vert\ne\varnothing$ and, $\forall s'\in\mathcal B(s), f(s',\Vert\phi\Vert)\subseteq\Vert\psi\Vert$. Consider the following full-domain extension $\diamond$ of $\circ$:
\begin{equation}
\label{circ_to_diamond}
K_s\diamond\phi=\left\{
\begin{array}{ll}K_s\circ\phi&\text{if }\Vert\phi\Vert\ne\varnothing\\
Cn(\phi)&\text{if }\Vert\phi\Vert=\varnothing.\end{array} \right.
\end{equation}
We want to show that the function defined in \eqref{circ_to_diamond} is a belief update function (Definition \ref{DEF:update}), that is, that it satisfies axioms $(K\diamond0)$-$(K\diamond7)$.\\[3pt]
$\bullet\,\,(K\diamond0)$. We need to show that $K_s\diamond\phi =Cn\left(K_s\diamond\phi\right)$. If $\Vert\phi\Vert\ne\varnothing$ then this follows from (B) of Lemma \ref{LEM:KfromModel}. If $\Vert\phi\Vert=\varnothing$ then it follows from the fact that $Cn (\phi)=Cn\left(Cn(\phi)\right)$ .\\[4pt]
$\bullet\,\,(K\diamond1)$. We need to show that $\phi\in K_s\diamond\phi$. If $\Vert\phi\Vert\ne\varnothing$ then this follows from (2) of Proposition \ref{PROP:validaxioms}. If $\Vert\phi\Vert=\varnothing$ then it follows from the fact that $\phi\in Cn(\phi)$.\\[4pt]
$\bullet\,\,(K\diamond2)$. We need to show that if $\phi\in K_s$ then $K_s\diamond\phi =K_s$. Assume that $\phi\in K_s$; then $\mathcal B(s)\subseteq\Vert\phi\Vert$ and thus, since by definition of frame (Definition \ref{DEF:frame}), $\mathcal B(s)\ne\varnothing$, $\Vert\phi\Vert\ne\varnothing$. Hence, by \eqref{circ_to_diamond}, $K_s\diamond\phi=K_s\circ\phi$ and the desired property follows from Proposition \ref{PROP:K_circ_2}.\\[4pt]
$\bullet\,\,(K\diamond3)$. We need to show that $K_s\diamond\phi=\Phi_0$ if and only if $\phi$ is a contradiction. If $\phi$ is a contradiction then $Cn(\phi)=\Phi_0$; furthermore, $\Vert\phi\Vert=\varnothing$ and thus, by \eqref{circ_to_diamond}, $K_s\diamond\phi=Cn(\phi)$. If $\phi$ is consistent and $\Vert\phi\Vert\ne\varnothing$, then, by (B) of Lemma \ref{LEM:KfromModel}, $K_s\circ\phi$ is consistent and thus $K_s\circ\phi\ne\Phi_0$ and, by \eqref{circ_to_diamond}, $K_s\diamond\phi=K_s\circ\phi$. Finally, if $\phi$ is consistent and $\Vert\phi\Vert=\varnothing$, then, by \eqref{circ_to_diamond}, $K_s\diamond\phi=Cn(\phi)$ and, since $\phi$ is consistent, $Cn(\phi)\ne\Phi_0$.\\[4pt]
$\bullet\,\,(K\diamond4)$. We need to show that if $\vdash(\phi\leftrightarrow\psi)$ then $K_s\diamond\phi=K_s\diamond\psi$. Assume that $\vdash(\phi\leftrightarrow\psi)$. If $\Vert\phi\Vert\ne\varnothing$ then this follows from (3) of Proposition \ref{PROP:validaxioms}. If $\Vert\phi\Vert=\varnothing$, then (since $\vdash(\phi\leftrightarrow\psi)$ implies that $\Vert\phi\Vert=\Vert\psi\Vert$) $\Vert\psi\Vert=\varnothing$ and thus $K_s\diamond\phi=Cn(\phi)$ and $K_s\diamond\psi=Cn(\psi)$ and, since $\vdash(\phi\leftrightarrow\psi)$, $Cn(\phi)=Cn(\psi)$.\\[4pt]
$\bullet\,\,(K\diamond5)$. We need to show that $K_s\diamond(\phi\wedge\psi)\subseteq(K_s\diamond\phi)+\psi$. If $\Vert\phi\Vert=\varnothing$ then $\Vert\phi\wedge\psi\Vert=\varnothing$ and, by \eqref{circ_to_diamond}, $K_s\diamond\phi=Cn(\phi)$ -- so that $(K_s\diamond\phi)+\psi=Cn\left(Cn(\phi)\cup\{\psi\} \right)$ -- and $K_s\diamond(\phi\wedge\psi)=Cn(\phi\wedge\psi)$. Fix an arbitrary $\chi\in Cn(\phi\wedge\psi)$. Then $\vdash(\phi\wedge\psi)\rightarrow\chi$ which implies that $\vdash\phi\rightarrow(\psi\rightarrow\chi)$, from which it follows that $\phi\rightarrow(\psi\rightarrow\chi)\in Cn(\phi)$ and thus $(\psi\rightarrow\chi)\in Cn(\phi)$ which, in turn, is equivalent to $\chi\in Cn\left(Cn(\phi)\cup\{\psi\} \right)$. Hence $Cn(\phi\wedge\psi)\subseteq Cn\left(Cn(\phi)\cup\{\psi\} \right)$. If $\Vert\phi\wedge\psi\Vert\ne\varnothing$ then $\Vert\phi\Vert\ne\varnothing$ so that, by \eqref{circ_to_diamond}, $K_s\diamond\phi=K_s\circ\phi$ and $K_s\diamond(\phi\wedge\psi)=K_s\circ(\phi\wedge\psi)$ and the desired property follows from Proposition \ref{PROP:KM5/AGM7}. Finally, if $\Vert\phi\Vert\ne\varnothing$ and $\Vert\phi\wedge\psi\Vert=\varnothing$ then $K_s\diamond\phi=K_s\circ\phi$ and $K_s\diamond(\phi\wedge\psi)=Cn(\phi\wedge\psi)$. Fix an arbitrary $\chi\in Cn(\phi\wedge\psi)$. Then $\vdash(\phi\wedge\psi)\rightarrow\chi$, from which it follows that $\vdash\phi\rightarrow(\psi\rightarrow\chi)$, so that since, by (B) of Lemma \ref{LEM:KfromModel}, $K_s\circ\phi$ is deductively closed, $(\phi\rightarrow(\psi\rightarrow\chi))\in K_s\circ\phi$. From this and the fact that, by (2) of Proposition \ref{PROP:validaxioms}, $\phi\in K_s\circ\phi$, it follows that $(\psi\rightarrow\chi)\in K_s\circ\phi$ which is equivalent to $\chi\in Cn((K_s\circ\phi)\cup\{\psi\})$. \hfill $\Box$\\[4pt]
$\bullet\,\,(K\diamond6)$. We need to show that if $\psi\in K_s\diamond\phi$ and $\phi\in K_s\diamond\psi$ then $K_s\diamond\phi=K_s\diamond\psi$. Assume that $\psi\in K_s\diamond\phi$ and $\phi\in K_s\diamond\psi$. If $\Vert\phi\Vert\ne\varnothing$, then $K_s\diamond\phi= K_s\circ\phi$; furthermore, since - by hypothesis - $\psi\in K_s\circ\phi$, $\forall s'\in\mathcal B(s), f(s',\Vert\phi\Vert)\subseteq\Vert\psi\Vert)$ which implies that $\Vert\psi\Vert\ne\varnothing$ because, by definition of frame (Definition \ref{DEF:frame}), $f(s',\Vert\phi\Vert)\ne\varnothing$. A similar argument shows that if $\Vert\psi\Vert\ne\varnothing$ then $\Vert\phi\Vert\ne\varnothing$. Thus there are only two cases to consider: (1) $\Vert\phi\Vert=\Vert\psi\Vert=\varnothing$ and (2) $\Vert\phi\Vert\ne\varnothing$ and $\Vert\psi\Vert\ne\varnothing$. Consider first the case where $\Vert\phi\Vert=\Vert\psi\Vert=\varnothing$. Then, by \eqref{circ_to_diamond}, $K_s\diamond\phi=Cn(\phi)$ and $K_s\diamond\psi=Cn(\psi)$. Then, since $\psi\in Cn(\phi)$, $\vdash(\phi\rightarrow\psi)$ and, since $\phi\in Cn(\psi)$, $\vdash(\psi\rightarrow\phi)$. Thus $\vdash(\phi\leftrightarrow\psi)$ and therefore $Cn(\phi)=Cn(\psi)$. Next consider the case where $\Vert\phi\Vert\ne\varnothing$ and $\Vert\psi\Vert\ne\varnothing$. In this case $K_s\diamond\phi=K_s\circ\phi$ and $K_s\diamond\psi=K_s\circ\psi$ and the desired property follows from Proposition \ref{PROP:KM6}.\\[4pt]
$\bullet\,\,(K\diamond7)$. We need to show that if $K_s$ is complete then $(K_s\diamond\phi)\cap(K_s\diamond\psi)\subseteq K_s\diamond(\phi\wedge\psi)$. If $(K_s\diamond\phi)\cap(K_s\diamond\psi)=\varnothing$ there is nothing to prove. Suppose, therefore, that $(K_s\diamond\phi)\cap(K_s\diamond\psi)\ne\varnothing$ and let $\chi\in(K_s\diamond\phi)\cap(K_s\diamond\psi)$. If $\Vert\phi\Vert\ne\varnothing$ and $\Vert\psi\Vert\ne\varnothing$, then $K_s\diamond\phi=K_s\circ\phi$ and $K_s\diamond\psi=K_s\circ\psi$ and the desired property follows from Proposition \ref{PROP:KM7}. If $\Vert\phi\Vert=\Vert\psi\Vert=\varnothing$ then $\Vert\phi\wedge\psi\Vert=\varnothing$ and thus $K_s\diamond\phi=Cn(\phi)$, $K_s\diamond\psi=Cn(\psi)$ and $K_s\diamond(\phi\wedge\psi)=Cn(\phi\wedge\psi)$. Since $\chi\in(K_s\diamond\phi)\cap(K_s\diamond\psi)$ it follows that $\vdash(\phi\rightarrow\chi)$ and $\vdash(\psi\rightarrow\chi)$ so that $\vdash((\phi\wedge\psi)\rightarrow\chi)$ and thus $\chi\in Cn(\phi\wedge\psi)$. If $\Vert\phi\Vert\ne\varnothing$ and $\Vert\psi\Vert=\varnothing$ then $\Vert\phi\wedge\psi\Vert=\varnothing$ and thus $K_s\diamond\phi=K_s\circ\phi$, $K_s\diamond\psi=Cn(\psi)$ and $K_s\diamond(\phi\wedge\psi)=Cn(\phi\wedge\psi)$. Since $\chi\in(K_s\diamond\psi)=Cn(\psi)$, $\vdash(\psi\rightarrow\chi)$  so that $\vdash((\phi\wedge\psi)\rightarrow\chi)$ and thus $\chi\in Cn(\phi\wedge\psi)$. The case where $\Vert\phi\Vert=\varnothing$ and $\Vert\psi\Vert\ne\varnothing$ is similar: $\vdash(\phi\rightarrow\chi)$ and thus $\vdash((\phi\wedge\psi)\rightarrow\chi)$ and $\chi\in Cn(\phi\wedge\psi)$. \\[6pt]
\textbf{(B)} Next we prove that, for every belief update function $\diamond$, there is a model based on an update frame such that the belief change function $\circ$ obtained at a state in that model coincides with $\diamond$ on the domain of $\circ$ and, furthermore, the domain of $\circ$ is the set of consistent formulas.  Once again, \emph{the purpose here is not to define the most natural model but to show that such a model exists}. Thus we will construct the simplest model.\\
Let $K\subset\Phi_0$ be consistent and deductively closed and let $\diamond:\Phi_0\to 2^{\Phi_0}$ be a belief update function based on $K$ (Definition \ref{DEF:update}). Define the following model $\left\langle {S,\mathcal B,f,V} \right\rangle$:
\begin{enumerate}
  \item $S$ is the set of maximally consistent sets (MCS) of formulas in $\Phi_0$.
  \item The valuation $V:\texttt{At}\rightarrow S$ is defined by $V(p)=\{s\in S: p\in s\}$, so that, for every $\phi\in\Phi_0$, the truth set of $\phi$, which - in this context - we denote by $\llbracket\phi\rrbracket$ instead of $\Vert\phi\Vert$, is $\llbracket\phi\rrbracket=\{s\in S: \phi\in s\}$.  If $\Psi\subseteq\Phi_0$, define $\llbracket\Psi\rrbracket=\{s\in S: \forall \phi\in \Psi, \phi\in s\}$. Note that $\llbracket\Psi\rrbracket\ne\varnothing$ if and only if $\Psi$ is consistent.
  \item For every $s\in S$, define $\mathcal B(s)=\llbracket K\rrbracket$.
  \item In order to define the selection function $f$, note first that $\llbracket\phi\rrbracket\ne\varnothing$ if and only if $\phi$ is consistent. Thus we only need to define $f(s,\Vert\phi\Vert)$ for $\phi$ consistent. Let $\Phi_{cn}\subseteq\Phi_0$ be the set of consistent formulas and let $\mathcal E = \{E\subseteq S:E=\llbracket\phi\rrbracket \text{ for some  } \phi\in\Phi_{cn}\}$. Define $f:\llbracket K\rrbracket\times \mathcal E\rightarrow 2^S$  as follows:
\begin{equation}
\label{EQ:f_for_B}
     f(s,\llbracket\phi\rrbracket) = \llbracket K\diamond\phi \rrbracket.
\end{equation}
\end{enumerate}
First we show that the frame so defined is an update frame (Definition \ref{DEF:update_frame}).\\[3pt]
$\bullet\,\,f(s,\llbracket\phi\rrbracket)\ne\varnothing$. This follows from axiom $(K\diamond3)$ since we are restricting attentions to $\phi\in\Phi_{cn}$.\\[3pt]
$\bullet\,\,f(s,\llbracket\phi\rrbracket)\subseteq\llbracket\phi\rrbracket$. This follows from axiom  $(K\diamond1)$ (since $\phi\in K\diamond\phi$, the set of MCS that satisfy all the formulas in $ K\diamond\phi$ is a subset of the of MCS that satisfy $\phi$).\\[3pt]
To prove Weak Centering we first prove the following lemma.
\begin{lemma}
\label{LEM:WCentering}
$\forall \phi\in\Phi_{cn},\,\,\llbracket K\rrbracket\cap\llbracket\phi\rrbracket=\llbracket Cn(K\cup\{\phi\})\rrbracket.$
\end{lemma}
\begin{proof}
By hypothesis, $K$ is deductively closed. Thus, $\forall \chi\in\Phi_0$,
\begin{equation}
\label{EQ:lem[Ku]PL_1}
\chi\in Cn(K\cup\{\phi\})\text{ if and only if } (\phi\rightarrow\chi)\in K.
\end{equation}
First we show that
$$\llbracket K\rrbracket\cap\llbracket\phi\rrbracket \subseteq\llbracket Cn(K\cup\{\phi\})\rrbracket.$$
Fix an arbitrary $s\in\llbracket K\rrbracket\cap\llbracket\phi\rrbracket$; we need to show that  $s\in\llbracket Cn(K\cup\{\phi\})\rrbracket$, that is, that, $\forall\chi\in Cn(K\cup\{\phi\}),\, \chi\in s$. Since $s\in\llbracket\phi\rrbracket$, $\phi\in s$. Fix an arbitrary $\chi\in Cn(K\cup\{\phi\})$; then, by \eqref{EQ:lem[Ku]PL_1}, $(\phi\rightarrow\chi)\in K$; thus, since $s\in\llbracket K\rrbracket$, $(\phi\rightarrow\chi)\in s$. Hence, since both $\phi$ and $\phi\rightarrow\chi$ belong to $s$ and $s$ is deductively closed (every MCS is deductively closed), $\chi\in s$.\\
Next we show that
$$\llbracket Cn(K\cup\{\phi\})\rrbracket\subseteq\llbracket K\rrbracket\cap\llbracket\phi\rrbracket.$$
Let $s\in \llbracket Cn(K\cup\{\phi\})\rrbracket$. Then, since $\phi\in Cn(K\cup\{\phi\})$, $\phi\in s$, that is, $s\in\llbracket\phi\rrbracket$. It remains to show that $s\in\llbracket K\rrbracket$, that is, that, for every $\chi\in K$, $\chi\in s$. Fix an arbitrary  $\chi\in K$; then, since, by hypothesis, $K$ is deductively closed, $(\phi\rightarrow\chi)\in K$. Thus, by \eqref{EQ:lem[Ku]PL_1}, $\chi\in Cn(K\cup\{\phi\})$ and thus, since $s\in\llbracket Cn(K\cup\{\phi\})\rrbracket$, $\chi\in s$.
\end{proof}
\noindent
$\bullet\,\,\text{if } s\in\llbracket\phi\rrbracket\text{ then }s\in f(s,\llbracket\phi\rrbracket)$. Let $s\in\llbracket K\rrbracket$ and $\phi\in\Phi_{cn}$. Assume that $s\in\Vert\phi\Vert$; then $s\in\llbracket K\rrbracket\cap\llbracket\phi\rrbracket$ so that, by Lemma \ref{LEM:WCentering}, $s\in\Vert Cn(K\cup\{\phi\})\Vert$. By Lemma \ref{LEM:*3forKM}, $K\diamond\phi\subseteq Cn(K\cup\{\phi\})$ from which it follows that $\llbracket Cn(K\cup\{\phi\}\rrbracket\subseteq\llbracket K\diamond\phi\rrbracket$. Hence $s\in \llbracket K\diamond\phi\rrbracket$ and, by \eqref{EQ:f_for_B}, $ \llbracket K\diamond\phi\rrbracket=f(s,\llbracket\phi\rrbracket).$\\[3pt]
$\bullet\,\, \text{ if } \mathcal B(s)\subseteq \llbracket\phi\rrbracket\text{ then, }\forall s'\in\mathcal B(s), f(s', \llbracket\phi\rrbracket)\subseteq\mathcal B(s).$ Fix an arbitrary $\phi\in\Phi_{cn}$ and an arbitrary MCS $s$ and recall that, by construction, $\mathcal B(s)=\llbracket K\rrbracket$ and, $\forall s'\in\mathcal B(s)$, $f(s',\llbracket\phi\rrbracket)=\llbracket K\diamond\phi\rrbracket$. Thus we need to show that if $\llbracket K\rrbracket\subseteq \llbracket\phi\rrbracket$ then $\llbracket K\diamond\phi\rrbracket\subseteq\llbracket K\rrbracket$. Assume that $\llbracket K\rrbracket\subseteq \llbracket\phi\rrbracket$, which is equivalent to $\phi\in K$. Then, by $(K\diamond2)$, $K\diamond\phi=K$ and thus  $\llbracket K\diamond\phi\rrbracket=\llbracket K\rrbracket$.\\[3pt]
In order to prove the next property we need the following lemma, which is similar to Lemma \ref{LEM:WCentering}.
\begin{lemma}
\label{4PKM5_2}
$\forall \phi,\psi\in\Phi_{cn},\,\,\llbracket K\diamond\phi\rrbracket\cap\llbracket\psi\rrbracket=\llbracket Cn((K\diamond\phi)\cup\{\psi\})\rrbracket.$
\end{lemma}
\begin{proof}
By $(K\diamond0)$, $K\diamond\phi$ is deductively closed. Thus, $\forall \chi\in\Phi_0$,
\begin{equation}
\label{EQ:lem4PKM5_2}
\chi\in Cn((K\diamond\phi)\cup\{\psi\})\text{ if and only if } (\psi\rightarrow\chi)\in K\diamond\phi.
\end{equation}
First we show that
$$\llbracket K\diamond\phi\rrbracket\cap\llbracket\psi\rrbracket \subseteq\llbracket Cn((K\diamond\phi)\cup\{\psi\})\rrbracket.$$
Fix an arbitrary $s\in\llbracket K\diamond\phi\rrbracket\cap\llbracket\phi\rrbracket$; we need to show that  $s\in\llbracket Cn((K\diamond\phi)\cup\{\psi\})\rrbracket$, that is, that, $\forall\chi\in Cn((K\diamond\phi)\cup\{\psi\}),\, \chi\in s$. Since $s\in\llbracket\psi\rrbracket$, $\psi\in s$. Fix an arbitrary $\chi\in Cn((K\diamond\phi)\cup\{\psi\})$; then, by \eqref{EQ:4PKM6}, $(\psi\rightarrow\chi)\in K\diamond\phi$; thus, since $s\in\llbracket K\diamond\phi\rrbracket$, $(\psi\rightarrow\chi)\in s$. Hence, since both $\psi$ and $\psi\rightarrow\chi$ belong to $s$ and $s$ is deductively closed, $\chi\in s$.\\
Next we show that
$$\llbracket Cn((K\diamond\phi)\cup\{\psi\})\rrbracket\subseteq\llbracket K\diamond\phi\rrbracket\cap\llbracket\psi\rrbracket.$$
Let $s\in \llbracket Cn((K\diamond\phi)\cup\{\psi\})\rrbracket$. Then, since $\psi\in Cn((K\diamond\phi)\cup\{\psi\})$, $\psi\in s$, that is, $s\in\llbracket\psi\rrbracket$. It remains to show that $s\in\llbracket K\diamond\phi\rrbracket$, that is, that, for every $\chi\in K\diamond\phi$, $\chi\in s$. Fix an arbitrary  $\chi\in K\diamond\phi$; then, since, by axiom $(K\diamond0)$, $K\diamond\phi$ is deductively closed, $(\psi\rightarrow\chi)\in K\diamond\phi$. Thus, by \eqref{EQ:4PKM6}, $\chi\in Cn((K\diamond\phi)\cup\{\psi\})$ and thus, since $s\in\llbracket Cn((K\diamond\phi)\cup\{\psi\})\rrbracket$, $\chi\in s$.
\end{proof}
\noindent
$\bullet\,\, \text{If, } \forall s'\in \mathcal B(s), \,f(s',\llbracket\phi\rrbracket\cap \llbracket\psi\rrbracket)\subseteq \llbracket\chi\rrbracket, \text{ then, }\forall s'\in \mathcal B(s),  \,f(s',\llbracket\phi\rrbracket)\,\cap\, \llbracket\psi\rrbracket\subseteq \llbracket\chi\rrbracket.$  Fix arbitrary $\phi,\psi,\chi\in\Phi_{cn}$ and an arbitrary MCS $s$ and recall that, by construction, $\mathcal B(s)=\llbracket K\rrbracket$,  $f(s',\llbracket\phi\rrbracket)=\llbracket K\diamond\phi\rrbracket$ and (since $\llbracket\phi\rrbracket\cap\llbracket\psi\rrbracket=\llbracket\phi\wedge\psi\rrbracket$) $f(s',\llbracket\phi\rrbracket\cap\llbracket\psi\rrbracket)=\llbracket K\diamond(\phi\wedge\psi)\rrbracket$. Thus it is sufficient to show that $\llbracket K\diamond\phi\rrbracket\cap\llbracket\psi\rrbracket\subseteq\llbracket K\diamond(\phi\wedge\psi)\rrbracket$. By axiom $(K\diamond5)$, $K\diamond(\phi\wedge\psi)\subseteq Cn((K\diamond\phi)\cup\{\psi\})$ from which it follows that
\begin{equation}
\label{EQ:lem4PKM5_3}
\llbracket Cn((K\diamond\phi)\cup\{\psi\})\rrbracket\subseteq\llbracket K\diamond(\phi\wedge\psi)\rrbracket.
\end{equation}
By Lemma \ref{4PKM5_2}, $\llbracket K\diamond\phi\rrbracket\cap\llbracket\psi\rrbracket=\llbracket Cn((K\diamond\phi)\cup\{\psi\})\rrbracket$. It follows from this and \eqref{EQ:lem4PKM5_3} that $\llbracket K\diamond\phi\rrbracket\cap\llbracket\psi\rrbracket\subseteq\llbracket K\diamond(\phi\wedge\psi)\rrbracket$.\\[3pt]
$\bullet\,\, \text{If, }\forall s'\in\mathcal B(s),\,  f(s',\llbracket\phi\rrbracket)\subseteq \llbracket\psi\rrbracket \text{ and }\, f(s',\llbracket\psi\rrbracket)\subseteq \llbracket\phi\rrbracket, \text{ then }\bigcup\limits_{s'\in\mathcal B(s)} {f(s',\llbracket\phi\rrbracket)} =\bigcup\limits_{s'\in\mathcal B(s)} {f(s',\llbracket\psi\rrbracket)}$. By \eqref{EQ:f_for_B} this reduces to:
\begin{equation}
\label{EQ:4PKM6}
\text{if } \llbracket K\diamond\phi\rrbracket\subseteq\llbracket \psi\rrbracket\text{ and }\llbracket K\diamond\psi\rrbracket\subseteq\llbracket \phi\rrbracket\text{ then }\llbracket K\diamond\phi\rrbracket=\llbracket K\diamond\psi\rrbracket.
\end{equation}
By $(K\diamond6)$, if $\psi\in K\diamond\phi$ and $\phi\in K\diamond\psi$ then $K\diamond\phi=K\diamond\psi$. Since $\psi\in K\diamond\phi$ if and only if $\llbracket K\diamond\phi\rrbracket\subseteq\llbracket \psi\rrbracket$, and $\phi\in K\diamond\psi$ if and only if $\llbracket K\diamond\psi\rrbracket\subseteq\llbracket \phi\rrbracket$, \eqref{EQ:4PKM6} follows directly from $(K\diamond6)$.\\[3pt]
$\bullet\,\, \text{If }\mathcal B(s) =\{s'\}\text{ then } f(s',\llbracket \phi\rrbracket\cup\llbracket \psi\rrbracket)\subseteq f(s',\llbracket \phi\rrbracket)\cup f(s',\llbracket \psi\rrbracket)$. First of all, note that $\llbracket\phi\rrbracket\cup\llbracket \psi\rrbracket=\llbracket\phi\vee\psi\rrbracket$. By $(K\diamond7)$, since $\llbracket K \rrbracket$ is a singleton (and thus $K$ is complete), $(K\diamond\phi)\cap(K\diamond\psi)\subseteq K\diamond(\phi\vee\psi)$, which implies that $\llbracket K\diamond(\phi\vee\psi)\rrbracket\subseteq\llbracket K\diamond\phi\rrbracket\cap\llbracket K\diamond\psi\rrbracket\subseteq\llbracket K\diamond\phi\rrbracket\cup\llbracket K\diamond\psi\rrbracket$. Finally, by \eqref{EQ:f_for_B}, $f(s',\llbracket \phi\rrbracket)=\llbracket K\diamond\phi\rrbracket$, $f(s',\llbracket \psi\rrbracket)=\llbracket K\diamond\psi\rrbracket$ and $f(s',\llbracket \phi\vee \psi\rrbracket)=\llbracket K\diamond(\phi\vee\psi)\rrbracket$.\\[6pt]
So far we have shown that the model that we have constructed is based on an update frame (Definition \ref{DEF:update_frame}). It remains to show that the belief update function $\diamond$ that we started with coincides with the partial belief change function $\circ$ defined by \eqref{RI} relative to some MCS $s$. Since, in the model that we constructed, for any two MCS's $s$ and $s'$, $\mathcal B(s)=\mathcal B(s')$, we can take an arbitrary MCS, call it $\hat s$. Let $K_{\hat s}=\{\phi\in\Phi_0:\mathcal B(\hat s)\subset \llbracket\phi\rrbracket\}$. By construction, $\mathcal B(\hat s)=\llbracket K\rrbracket$ and thus $K_{\hat s}=K$. Let $\circ$ be the belief change function based on $K_{\hat s}=K$ defined by \eqref{RI}. We need to show that (since the domain of $\circ$ is $\Phi_{cn}$), $\forall\phi\in\Phi_{cn}, K\diamond\phi=K\circ\phi$. First we show that $K\diamond\phi\subseteq K\circ\phi$. Let $\psi\in K\diamond\phi$; then $\psi\in s$ for all $s\in\llbracket K\diamond\phi\rrbracket$, that is, $\llbracket K\diamond\phi\rrbracket\subseteq\llbracket\psi\rrbracket$. By \eqref{EQ:f_for_B}, $\forall s\in\mathcal B(\hat s)$, $f(s,\llbracket\phi\rrbracket)=\llbracket K\diamond\phi\rrbracket$. Thus, $\forall s\in\mathcal B(\hat s)$, $f(s,\llbracket\phi\rrbracket)\subseteq\llbracket\psi\rrbracket$, that is, $\psi\in K\circ\phi$. Next we show that $K\circ\phi\subseteq K\diamond\phi$. Let $\psi\in K\circ\phi$, that is, $\forall s\in\mathcal B(\hat s), f(s,\llbracket\phi\rrbracket)\subseteq\llbracket\psi\rrbracket$. By \eqref{EQ:f_for_B}, $\forall s\in\mathcal B(\hat s)$, $f(s,\llbracket\phi\rrbracket)=\llbracket K\diamond\phi\rrbracket$. Thus $\llbracket K\diamond\phi\rrbracket\subseteq\llbracket\psi\rrbracket$, that is, $\forall s\in\llbracket K\diamond\phi\rrbracket$, $\psi\in s$. Hence $\psi\in K\diamond\phi$.\hfill$\square$\\[6pt]
\noindent
\textbf{Proof of Proposition \ref{PROP:char_revision}.} Recall that $\vdash\phi$ means that $\phi$ is a tautology.\\
 \textbf{(A)} We need to show that the partial belief change function obtained at a state of an arbitrary model based on a revision frame, can be extended to a full-domain AGM belief revision function. We note again that the purpose here is not to define the most natural extension, but to show that such an extension is in fact possible.\\
 Fix an arbitrary frame $F\in\mathcal F_{\mathlarger{\ast}}$, an arbitrary model based on $F$,  an arbitrary state $s$ and let $K_s=\{\phi\in\Phi_0:\mathcal B(s)\subseteq\Vert\phi\Vert\}$. Let $\circ$ be the belief change function based on $K_s$ defined by \eqref{RI}, that is, $\psi\in K_s\circ\phi$ if and only if, $\forall s'\in\mathcal B(s), f(s',\Vert\phi\Vert)\subseteq\Vert\psi\Vert$. Consider the following full-domain extension $\ast$ of $\circ$:
\begin{equation}
\label{circ_to_ast}
K_s\ast\phi=\left\{
\begin{array}{ll}K_s\circ\phi&\text{if }\Vert\phi\Vert\ne\varnothing\\
Cn(\phi)&\text{if }\Vert\phi\Vert=\varnothing.\end{array} \right.
\end{equation}
We want to show that the function defined in \eqref{circ_to_ast} satisfies axioms $(K\ast1)$-$(K\ast8)$.\\[3pt]
$\bullet\,\,(K\ast1)$. We need to show that $K_s\ast\phi =Cn\left(K_s\ast\phi\right)$. If $\Vert\phi\Vert\ne\varnothing$ then this follows from (B) of Lemma \ref{LEM:KfromModel}. If $\Vert\phi\Vert=\varnothing$ then it follows from the fact that $Cn (\phi)=Cn\left(Cn(\phi)\right)$ .\\[4pt]
$\bullet\,\,(K\ast2)$. We need to show that $\phi\in K_s\ast\phi$. If $\Vert\phi\Vert\ne\varnothing$ then this follows from (2) of Proposition \ref{PROP:validaxioms}. If $\Vert\phi\Vert=\varnothing$ then it follows from the fact that $\phi\in Cn(\phi)$.\\[4pt]
$\bullet\,\,(K\ast3)$. We need to show that $K_s\ast\phi\subseteq Cn(K_s\cup\{\phi\})$. If $\Vert\phi\Vert\ne\varnothing$ then this follows from (1) of Proposition \ref{PROP:validaxioms}. If $\Vert\phi\Vert=\varnothing$ then $\Vert\lnot\phi\Vert=S$ and thus $\mathcal B(s)\subseteq\Vert\lnot\phi\Vert$, that is,  $\lnot\phi\in K_s$ which implies that $Cn(K_s\cup\{\phi\})=\Phi_0.$\\[4pt]
$\bullet\,\,(K\ast4)$. We need to show that if $\lnot\phi\notin K_s$ then $K_s\subseteq K_s\ast\phi$. Since $\lnot\phi\notin K_s$, $\mathcal B(s)\cap\Vert\phi\Vert\ne\varnothing$ so that $\Vert\phi\Vert\ne\varnothing$ and thus, by Proposition \ref{PROP:AGM4}, $K_s\subseteq K_s\ast\phi$.\\[4pt]
$\bullet\,\,(K\ast5)$. We need to show that  $K_s\ast\phi=\Phi_0$ if and only if $\vdash\lnot\phi$. If $\vdash\lnot\phi$ then $Cn(\phi)=\Phi_0$; furthermore, $\Vert\phi\Vert=\varnothing$ and thus, by \eqref{circ_to_ast}, $K_s\ast\phi=Cn(\phi)$. If $\phi$ is consistent and $\Vert\phi\Vert\ne\varnothing$, then, by (B) of Lemma \ref{LEM:KfromModel}, $K_s\circ\phi$ is consistent and thus $K_s\circ\phi\ne\Phi_0$ and, by \eqref{circ_to_ast}, $K\ast\phi=K\circ\phi$. Finally, if $\phi$ is consistent and $\Vert\phi\Vert=\varnothing$, then, by \eqref{circ_to_ast}, $K_s\ast\phi=Cn(\phi)$ and, since $\phi$ is consistent, $Cn(\phi)\ne\Phi_0$.\\[4pt]
$\bullet\,\,(K\ast6)$. We need to show that if $\vdash(\phi\leftrightarrow\psi)$ then $K_s\ast\phi=K_s\ast\psi$. Assume that $\vdash(\phi\leftrightarrow\psi)$. If $\Vert\phi\Vert\ne\varnothing$ then this follows from (3) of Proposition \ref{PROP:validaxioms}. If $\Vert\phi\Vert=\varnothing$, then (since $\vdash(\phi\leftrightarrow\psi)$ implies that $\Vert\phi\Vert=\Vert\psi\Vert$) $\Vert\psi\Vert=\varnothing$ and thus $K_s\ast\phi=Cn(\phi)$ and $K_s\ast\psi=Cn(\psi)$ and, since $\vdash(\phi\leftrightarrow\psi)$, $Cn(\phi)=Cn(\psi)$.\\[4pt]
$\bullet\,\,(K\ast7)$. We need to show that $K_s\ast(\phi\wedge\psi)\subseteq (K_s\ast\phi)+\psi$. If $\Vert\phi\Vert=\varnothing$ then $\Vert\phi\wedge\psi\Vert=\varnothing$ and, by \eqref{circ_to_ast}, $K_s\ast\phi=Cn(\phi)$ -- so that $(K_s\ast\phi)+\psi=Cn\left(Cn(\phi)\cup\{\psi\} \right)$ -- and $K_s\ast(\phi\wedge\psi)=Cn(\phi\wedge\psi)$. Fix an arbitrary $\chi\in Cn(\phi\wedge\psi)$. Then $\vdash(\phi\wedge\psi)\rightarrow\chi$ which implies that $\vdash\phi\rightarrow(\psi\rightarrow\chi)$, from which it follows that $\phi\rightarrow(\psi\rightarrow\chi)\in Cn(\phi)$ and thus $(\psi\rightarrow\chi)\in Cn(\phi)$ which, in turn, is equivalent to $\chi\in Cn\left(Cn(\phi)\cup\{\psi\} \right)$. Hence $Cn(\phi\wedge\psi)\subseteq Cn\left(Cn(\phi)\cup\{\psi\} \right)$. If $\Vert\phi\wedge\psi\Vert\ne\varnothing$ then $\Vert\phi\Vert\ne\varnothing$ so that, by \eqref{circ_to_ast}, $K_s\ast\phi=K_s\circ\phi$ and $K_s\ast(\phi\wedge\psi)=K_s\circ(\phi\wedge\psi)$ and the desired property follows from Proposition \ref{PROP:KM5/AGM7}. Finally, if $\Vert\phi\Vert\ne\varnothing$ and $\Vert\phi\wedge\psi\Vert=\varnothing$ then $K_s\ast\phi=K_s\circ\phi$ and $K_s\ast(\phi\wedge\psi)=Cn(\phi\wedge\psi)$. Fix an arbitrary $\chi\in Cn(\phi\wedge\psi)$. Then $\vdash(\phi\wedge\psi)\rightarrow\chi$, from which it follows that $\vdash\phi\rightarrow(\psi\rightarrow\chi)$, so that since, by (B) of Lemma \ref{LEM:KfromModel}, $K_s\circ\phi$ is deductively closed, $(\phi\rightarrow(\psi\rightarrow\chi))\in K_s\circ\phi$. From this and the fact that, by (2) of Proposition \ref{PROP:validaxioms}, $\phi\in K_s\circ\phi$, it follows that $(\psi\rightarrow\chi)\in K_s\circ\phi$ which is equivalent to $\chi\in Cn((K_s\circ\phi)\cup\{\psi\})$. \hfill $\Box$\\[4pt]
$\bullet\,\,(K\ast8)$. We need to show that if $\lnot\psi\notin K_s\ast\phi$ then $(K_s\ast\phi)+\psi\subseteq K_s\ast(\phi\wedge\psi)$. If $\Vert\phi\Vert=\varnothing$ then, as shown in the proof of $(K\ast7)$, $K\ast(\phi\wedge\psi)=(K\ast\phi)+\psi$. If $\Vert\phi\Vert\ne\varnothing$, then the result follows from Proposition \ref{PROP:AGM8}. \\[6pt]
\noindent
\textbf{(B)} Next we prove that, for every AGM belief revision function $\ast$, there is a model based on a revision frame such that the belief change function $\circ$ obtained at a state in that model coincides with $\ast$ on the domain of $\circ$ and, furthermore, the domain of $\circ$ is the set of consistent formulas.  Once again, the purpose here is not to define the most natural model but to show that such a model exists.\\
Let $K\subset\Phi_0$ be consistent and deductively closed and let $\ast:\Phi_0\to 2^{\Phi_0}$ be an AGM belief revision function based on $K$ (Definition \ref{DEF:revision}). Define the following model $\left\langle {S,\mathcal B,f,V} \right\rangle$:
\begin{enumerate}
  \item $S$ is the set of maximally consistent sets (MCS) of formulas in $\Phi_0$.
  \item The valuation $V:\texttt{At}\rightarrow S$ is defined by $V(p)=\{s\in S: p\in s\}$, so that, for every $\phi\in\Phi_0$, the truth set of $\phi$, which - in this context - we denote by $\llbracket\phi\rrbracket$ instead of $\Vert\phi\Vert$, is $\llbracket\phi\rrbracket=\{s\in S: \phi\in s\}$.  If $\Psi\subseteq\Phi_0$, define $\llbracket\Psi\rrbracket=\{s\in S: \forall \phi\in \Psi, \phi\in s\}$.
  \item For every $s\in S$ and define $\mathcal B(s)=\llbracket K\rrbracket$.
  \item In order to define the selection function $f$, note first that $\llbracket\phi\rrbracket\ne\varnothing$ if and only if $\phi$ is consistent. Thus we only need to define $f(s,\Vert\phi\Vert)$ for $\phi$ consistent. Let $\Phi_{cn}\subseteq\Phi_0$ be the set of consistent formulas and let $\mathcal E = \{E\subseteq S:E=\llbracket\phi\rrbracket \text{ for some  } \phi\in\Phi_{cn}\}$. Define $f:\llbracket K\rrbracket\times \mathcal E\rightarrow 2^S$  as follows:
\begin{equation}
\label{EQ:f_for_*B}
     f(s,\llbracket\phi\rrbracket) = \llbracket K\ast\phi \rrbracket.
\end{equation}
\end{enumerate}
First we show that the frame so defined is a revision frame (Definition \ref{DEF:revision-frame}).\\[3pt]
$\bullet\,\,f(s,\llbracket\phi\rrbracket)\ne\varnothing$. This follows from axiom $(K\ast5)$ since we are restricting attentions to $\phi\in\Phi_{cn}$.\\[3pt]
$\bullet\,\,f(s,\llbracket\phi\rrbracket)\subseteq\llbracket\phi\rrbracket$. This follows from axiom  $(K\ast2)$ (since $\phi\in K\ast\phi$, $\llbracket K\ast\phi\rrbracket\subseteq\llbracket\phi\rrbracket$ and, by \eqref{EQ:f_for_*B}, $f(s,\llbracket\phi\rrbracket)=\llbracket K\ast\phi\rrbracket$.\\[3pt]
$\bullet$ If $s\in\llbracket\phi\rrbracket$ then $s\in f(s,\llbracket\phi\rrbracket)$. Let $s\in\llbracket K\rrbracket$ and $\phi\in\Phi_{cn}$. Assume that $s\in\Vert\phi\Vert$; then $s\in\llbracket K\rrbracket\cap\llbracket\phi\rrbracket$ so that, by Lemma \ref{LEM:WCentering}, $s\in\Vert Cn(K\cup\{\phi\})\Vert$. By Lemma \ref{LEM:*3forKM}, $K\ast\phi\subseteq Cn(K\cup\{\phi\})$ from which it follows that $\llbracket Cn(K\cup\{\phi\}\rrbracket\subseteq\llbracket K\ast\phi\rrbracket$. Hence $s\in \llbracket K\ast\phi\rrbracket=f(s,\llbracket\phi\rrbracket).$\\[3pt]
$\bullet\,\, \text{ if } \mathcal B(s)\cap \llbracket\phi\rrbracket\ne\varnothing\text{ then, }\forall s'\in\mathcal B(s), f(s', \llbracket\phi\rrbracket)\subseteq\mathcal B(s)\cap \llbracket\phi\rrbracket.$ Fix an arbitrary $\phi\in\Phi_{cn}$ and an arbitrary MCS $s$ and recall that, by construction, $\mathcal B(s)=\llbracket K\rrbracket$ and, $\forall s'\in\mathcal B(s)$, $f(s',\llbracket\phi\rrbracket)=\llbracket K\ast\phi\rrbracket$. Thus we need to show that if $\llbracket K\rrbracket\cap \llbracket\phi\rrbracket\ne\varnothing$ then $\llbracket K\ast\phi\rrbracket\subseteq\llbracket K\rrbracket\cap \llbracket\phi\rrbracket$. Assume that $\llbracket K\rrbracket\cap \llbracket\phi\rrbracket\ne\varnothing$, which is equivalent to $\lnot\phi\notin K$. Then, by $(K\ast4)$, $K\subseteq K\ast\phi$ which implies that  $\llbracket K\ast\phi\rrbracket\subseteq\llbracket K\rrbracket$ and thus $\llbracket K\ast\phi\rrbracket\cap\llbracket\phi\rrbracket\subseteq\llbracket K\rrbracket\cap\llbracket\phi\rrbracket$. Furthermore, by $(K\ast 2)$, $\llbracket K\ast\phi\rrbracket\subseteq\llbracket\phi\rrbracket$, so that $\llbracket K\ast\phi\rrbracket=\llbracket K\ast\phi\rrbracket\cap\llbracket\phi\rrbracket$. \\[3pt]
$\bullet$ if, $\exists \hat s\in\mathcal B(s)$ such that $f(\hat s,E)\cap F\ne\varnothing$ then, $\forall s'\in\mathcal B(s),\, f(s',E\cap F)\,\,{\mathlarger \subseteq}$ $\bigcup\limits_{s''\in\mathcal B(s)} (f(s'',E)\cap F)$. Fix arbitrary $\phi,\psi,\chi\in\Phi_{cn}$ and an arbitrary MCS $s$ and recall that, by construction, $\mathcal B(s)=\llbracket K\rrbracket$, $\forall s'\in\mathcal B(s), f(s',\llbracket\phi\rrbracket)=\llbracket K\ast\phi\rrbracket$ and (since $\llbracket\phi\rrbracket\cap\llbracket\psi\rrbracket=\llbracket\phi\wedge\psi\rrbracket$) $f(s',\llbracket\phi\rrbracket\cap\llbracket\psi\rrbracket)=\llbracket K\ast(\phi\wedge\psi)\rrbracket$. Thus we need to show that if $\llbracket K\ast\phi\rrbracket\cap\llbracket\psi\rrbracket\ne\varnothing$ then $\llbracket K\ast(\phi\wedge\psi)\rrbracket\subseteq\llbracket K\ast\phi\rrbracket\cap\llbracket\psi\rrbracket$. Assume that $\llbracket K\ast\phi\rrbracket\cap\llbracket\psi\rrbracket\ne\varnothing$. Then $\llbracket K\ast\phi\rrbracket\nsubseteq\llbracket\lnot\psi\rrbracket$, that is, $\lnot\psi\notin K\ast\phi$. Hence, by $(K\ast8)$, $(K\ast\phi)+\psi\subseteq K\ast(\phi\wedge\psi)$, from which it follows that $\llbracket K\ast(\phi\wedge\psi)\rrbracket\subseteq\llbracket (K\ast\phi)+\psi\rrbracket$. Finally, by Lemma \ref{4PKM5_2},\footnote{
Modify the statement of Lemma \ref{4PKM5_2} by replacing $\diamond$ with $\ast$; then the same proof applies, provided that one appeals  to axiom $(K\ast1)$ instead of axiom $(K\diamond0)$.
}\ 
$\llbracket (K\ast\phi)+\psi\rrbracket=\llbracket K\ast\phi\rrbracket\cap\llbracket\psi\rrbracket$.\\[4pt]
So far we have shown that the model that we have constructed is based on a revision frame (Definition \ref{DEF:revision-frame}). It remains to show that the belief revision function $\ast$ that we started with coincides with the partial belief change function $\circ$ defined by \eqref{RI} relative to some MCS $s$. Since, in the model that we constructed, for any two MCS's $s$ and $s'$, $\mathcal B(s)=\mathcal B(s')$, we can take an arbitrary MCS, call it $\hat s$. Let $K_{\hat s}=\{\phi\in\Phi_0:\mathcal B(\hat s)\subset \llbracket\phi\rrbracket\}$. By construction, $\mathcal B(\hat s)=\llbracket K\rrbracket$ and thus $K_{\hat s}=K$. Let $\circ$ be the belief change function based on $K_{\hat s}=K$ defined by \eqref{RI}. We need to show that, $\forall\phi\in\Phi_{cn}, K\ast\phi=K\circ\phi$.\\
First we show that $K\ast\phi\subseteq K\circ\phi$. Let $\psi\in K\ast\phi$; then $\psi\in s$ for all $s\in\llbracket K\ast\phi\rrbracket$, that is, $\llbracket K\ast\phi\rrbracket\subseteq\llbracket\psi\rrbracket$. By \eqref{EQ:f_for_*B}, $\forall s\in\mathcal B(\hat s)$, $f(s,\llbracket\phi\rrbracket)=\llbracket K\ast\phi\rrbracket$. Thus, $\forall s\in\mathcal B(\hat s)$, $f(s,\llbracket\phi\rrbracket)\subseteq\llbracket\psi\rrbracket$, that is, $\psi\in K\circ\phi$.\\
Next we show that $K\circ\phi\subseteq K\ast\phi$. Let $\psi\in K\circ\phi$, that is, $\forall s\in\mathcal B(\hat s)$, $f(s,\llbracket\phi\rrbracket)\subseteq\llbracket\psi\rrbracket$. By \eqref{EQ:f_for_*B}, $\forall s\in\mathcal B(\hat s)$, $f(s,\llbracket\phi\rrbracket)=\llbracket K\ast\phi\rrbracket$. Thus $\llbracket K\ast\phi\rrbracket\subseteq\llbracket\psi\rrbracket$, that is, $\forall s\in\llbracket K\ast\phi\rrbracket$, $\psi\in s$. Hence $\psi\in K\ast\phi$.\hfill$\square$
\bibliographystyle{abbrv}

\end{document}